\documentclass[fleqn,usenatbib]{mnras}
\usepackage{amsmath}
\usepackage{newtxtext,newtxmath}

\usepackage[T1]{fontenc}

\DeclareRobustCommand{\VAN}[3]{#2}
\let\VANthebibliography\thebibliography
\def\thebibliography{\DeclareRobustCommand{\VAN}[3]{##3}\VANthebibliography}

\usepackage{hyperref}
\hypersetup{
  colorlinks=true,        
  linkcolor=blue,         
  citecolor=cyan,         
  urlcolor=cyan            
}
\usepackage{mathrsfs}

\usepackage{graphicx}	
\usepackage{amsmath}	

\newcommand{\FMONEA}{\texttt{FIL-M1}~}
\newcommand{\FMONEB}{\texttt{FIL-M1}}

\newcommand{\eg}{e.g.,~}
\newcommand{\ie}{i.e.,~}

\title[A practical guide to a moment approach in NR]{A practical guide to
  a moment approach for neutrino transport in numerical relativity}

\author[C. Musolino et al.]{
Carlo Musolino,$^{1}$\thanks{E-mail: musolno@itp.uni-frankfurt.de (KTS)}
Luciano Rezzolla$^{1,2,3}$
\\
$^{1}$Institut f\"ur Theoretische Physik, Goethe Universit\"at, Max-von-Laue-Str. 1, 60438 Frankfurt am Main, Germany\\
$^{2}$Frankfurt Institute for Advanced Studies, Ruth-Moufang-Str 1, 60438 Frankfurt am Main, Germany\\
$^{3}$School of Mathematics, Trinity College, Dublin 2, Ireland
}

\pubyear{2022}

\begin{document}
\label{firstpage}
\pagerange{\pageref{firstpage}--\pageref{lastpage}}
\maketitle

\begin{abstract}
The development of a neutrino moment based radiative-transfer code to
simulate binary neutron-star mergers can easily become an obstacle path
because of the numerous ways in which the solution of the equations may
fail. We describe the implementation of the grey M1 scheme in our fully
general-relativistic magnetohydrodynamics code and detail those choices
and strategies that could lead either to a robust scheme or to a series
of failures. In addition, we present new tests designed to show the
consistency and accuracy of our code in conditions that are similar to
realistic merging conditions and introduce a new, publicly available,
benchmark based on the head-on collision of two neutron stars. This test,
which is computationally less expensive than a complete merging binary
but has all the potential pitfalls of the full scenario, can be used to
compare future implementations of M1 schemes with the one presented here.
\end{abstract}




\section{Introduction}

The merger of binary neutron-star (BNS) system offers a window into some
of the most extreme physics in the universe \citep{Baiotti2016,
  Paschalidis2016}. They are linked to short gamma-ray bursts
\citep{Rezzolla:2011, Palenzuela2013a, Kiuchi2015}, to the synthesis of
heavy elements through rapid neutron capture
processes~\citep{Metzger:2010} and, of course, to the emission of
gravitational waves (GWs). As such, understanding in detail the dynamics
of these systems is a key challenge that the scientific community has
been addressing with increasing levels of sophistication for over a
decade. The first observation of GWs from a BNS merger by the LIGO-VIRGO
collaboration~\citep{Abbott2017}, together with a rich set of
counterparts in the electromagnetic spectrum~\citep{Abbott2017b,
  Abbott2017d}, has provided constraints on gravity \citep{Abbott2019b}
as well as precious information about the properties of matter at the
extreme density regimes realised in the cores of neutron
stars~\citep[see, \eg][]{Abbott2018b, Bauswein2017b, Margalit2017,
  Rezzolla2017, Most2018}. Future observations of GWs from late BNS
inspirals and early post-merger emission by the hypermassive neutron star
(HMNS) remnant by current and future-generation detectors
\citep{Punturo2010b} will further constrain the equation of state (EOS)
of matter above nuclear saturation density, thus probing a regime where
the properties of the QCD phase diagram cannot be easily explored by
standard techniques. However, GWs from the inspiral and post-merger of a
BNS system alone cannot provide us with the all the details of this
complex picture. Particularly important are the explanation and accurate
modelling of key phenomena associated with these events, such as the
launching of a relativistic jet~\citep{Rezzolla:2011, Just2016,
  Ciolfi2020, Hayashi2021}, the composition and properties of the ejecta
leading to the emission of a kilonova~\citep{Bovard2017, Papenfort2018,
  Combi2022}, and the long-timescale evolution of the system beyond the
gravity-dominated regime. All of these questions need to be addressed
both for their astrophysical interest as well as for the complementary
information they may provide about the EOS of neutron-star matter
\citep{Combi2022, Papenfort:2022ywx}.

Modelling the system self-consistently and over timescales that are
hundreds or thousand times the dynamical timescale proves to be a
formidable challenge both from the theoretical as well as from the
computational perspective: if the inspiral is mostly dictated by gravity
and the strong interactions between the constituent particles of the NS
fluid, considerably more complex physical phenomena become relevant
at the merger. First, it was shown that even small magnetic fields in the
two constituent NSs will amplify due to turbulence and Kelvin-Helmholtz
instabilities~\citep{Kiuchi2015a, Aguilera-Miret2021, Chabanov2022} to
reach almost equipartition with the fluid kinetic energy, making
electromagnetic forces extremely relevant on the long timescale in which
the remnant evolves. Moreover, weak interactions play a crucial role in
the dynamics of the system after merger for a number reasons. First, they
contribute to reprocessing the ejected material increasing its electron
fraction~\citep{Combi2022, Radice2016, Sekiguchi2016}. Second, they
provide a pressure contribution that will lead to the shedding of
additional reprocessed material in the form of a neutrino
wind~\citep{Dessart2009, Perego2014, Fujibayashi2017}. Third, on the
timescale of the diffusion of neutrinos through the optically-thick
remnant, they cool the system very efficiently, thus counteracting the
heating due to magnetic-field turbulence. Finally, it has long been
suggested that neutrinos may have a role in the powering of a collimated
polar outflow from the merger remnant~\citep{Eichler89, Ruffert97,
  Just2016}.

Several approaches are available in the literature for the inclusion of
the effects of neutrinos in general-relativistic hydrodynamical or
magnetohydrodynamical (GRMHD) simulations of BNS mergers. These range
from very simple and computationally efficient ``leakage-type''
schemes~\citep{Ruffert97, Galeazzi2013, Most2019b}, where the local
heating/cooling rates are directly estimated from the reaction
cross-sections corrected with a diffusion prescription, over to the
so-called ``moment schemes'', where a varying number of moments of the
Boltzmann equation for neutrinos is solved~\citep{Rezzolla1994,
  Foucart2015a, Just2015b, Kuroda2016, Skinner2019, MelonFuksman2019,
  Weih2020b, Radice2022, Sun:2022vri, Izquierdo2022}. The most advanced approaches
even consider the direct solution of the radiative transfer equation via
Montecarlo or other methods~\citep{Radice2013, Foucart2020montecarlo,
  Roth2022}.

In this paper we present \FMONEB: a new implementation of a two-moment
(M1) scheme for general-relativistic radiative transfer applied to BNS
mergers. Obviously, this is not the first paper discussing such an
implementation and papers with a similar spirit have been published very
recently~\citep[see, \eg][]{Weih2020b, Radice2022}. Here, however,
besides describing in detail our implementation of the grey M1 scheme in
a fully general-relativistic MHD, we also concentrate on highlighting the
possible pitfalls one may encounter in the process of developing a
similar code. In addition, we present new tests designed to show the
consistency and accuracy of our code in conditions that are similar to
realistic merging conditions: the head-on collision of two neutron
stars. The neutrino luminosities obtained in this way and the evolution
of the maximum temperature and density can serve therefore as an
effective benchmark against which test new implementations.

The content of the paper is organised as follows: in
Sec. ~\ref{sec:m1_theory} we review the theoretical background of
moment-based methods for radiative transfer and the basic equations
involved. In Sec. ~\ref{sec:numerics} we describe in detail the numerical
scheme implemented in our code, paying special attention to troublesome
aspects of the algorithm which are key to a successful application of the
methods. Finally, in Sec.~\ref{sec:implementation_tests} we present
results from standard implementation tests, as well as from a set of
realistic tests designed to probe the accuracy and convergence properties
of the code when presented with a realistic neutron-star simulation.

\section{M1 Scheme for General Relativistic radiation transfer}
\label{sec:m1_theory} 

In the limit where neutrinos are considered to be massless particles,
their evolution can be expressed in terms of the radiation intensity,
which in turn has to satisfy the Boltzmann equation~\citep[see,
  \eg][]{Rezzolla_book:2013}. Since this equation is intrinsically
$3+3+1$ dimensional, its direct solution is computationally too
expensive~\citep[see however][for some efficient methods to discretize
  the Boltzmann equation]{Weih2020c} to be feasible in conjunction with
the solution of the coupled Einstein and GRMHD system. Following the
moment formalism by Thorne~\citep{Thorne1981}, the M1 scheme consists in
evolving the zeroth and first moments of the radiation intensity whilst
providing an approximate closure for the second. In the following, we
will assume that spacetime is coupled with an ``ordinary'' perfect fluid
described by the equations of GRMHD which couples to neutrinos by weak
processes to be defined below. To this end, we consider the second moment
of the radiation transfer equation which reads~\citep[see ][ for a
  detailed derivation]{Shibata2011} 
\begin{equation} \label{eq:mom_spectral}
\nabla_\alpha M_{(f)}^{\beta \alpha} - \frac{\partial}{\partial f}
\left( f M_{(f)}^{\beta \alpha \gamma } \nabla_\gamma u_\beta \right)
= S_{(f)}^{\alpha}\,,
\end{equation}
where $M_{(f)}^{\alpha\beta}$ is the radiation intensity integrated
twice over the (null) direction of propagation, $u^{\alpha}$ is the fluid
four-velocity, $f$ is the frequency of the neutrino,
and $S_{(f)}^\alpha$ is the moment of the source
term. 

For simplicity, we will here consider a so-called ``grey scheme'', \ie
one where we evolve the moments integrated over the neutrino frequency
(energy) $f$. As a result, the second term in
Eq.~\eqref{eq:mom_spectral} vanishes out and we are left with a
conservation equation for the radiation energy momentum
\begin{equation}
  \label{eq:grey_moment}
  \nabla_\beta T^{\alpha \beta}_\text{rad} = \nabla_\beta \int_0^\infty
  df M_{(f)}^{\alpha \beta} =: S^\beta\,.
\end{equation}
Obviously, the total energy momentum tensor, which is the sum of the
energy-momentum tensor of the ordinary fluid $T^{\alpha
  \beta}_\text{flu}$ and of the radiation fluid $T^{\alpha
  \beta}_\text{rad}$ will have to be conserved, so that 
\begin{equation}
  \label{eq:mhd_cons}
  \nabla_\beta T^{\alpha \beta}_\text{flu}= - S^\beta = - \nabla_\beta T^{\alpha \beta}_\text{rad}\,.
\end{equation}
In the expression above, and for a magnetised fluid in the ideal-MHD
limit, the energy-momentum tensor reads 
\begin{equation}
  T^{\alpha \beta}_\text{flu} :=\left(\rho h + b^2\right) u^{\alpha}u^{\beta}+
  \left(p + \frac{b^2}{2} \right) g^{\alpha \beta} - b^{\alpha}b^{\beta}\,,
\end{equation}
where $\rho$ and $h$ are the rest-mass density and specific enthalpy,
$u^\alpha$ is the fluid's four-velocity, $g^{\alpha \beta}$ the four-metric 
and $b^{\alpha}$ the magnetic field four-vector in the frame comoving with the fluid, with
$b^2:=b^{\alpha}b_{\alpha}$. On the other hand, the energy-momentum
tensor of neutrinos can be expressed in terms of the first three moments
of the radiation intensity in a frame comoving with the fluid
\begin{equation}
  \label{eq:tmunu_rad_fluid}
  T^{\alpha \beta}_\text{rad} := J\, u^\alpha u^\beta + H^\alpha u^\beta +
  H^\beta u^\alpha + L^{\alpha\beta}\,.
\end{equation}
The M1 scheme has the advantage of providing immediate physical
interpretation of the quantities at hand: in analogy with the theory of
ideal fluids we identify $J$ as the energy density, $H^\alpha$ as the
energy flux density and $L^{\alpha\beta}$ as the pressure tensor of the
radiation as measured in the fluid comoving frame. It is, however, worth
noting that interpreting the neutrino field as a ideal fluid only makes
sense in completely optically-thick media where the radiation is in local
thermodynamic equilibrium with the background matter~\citep[see
  discussion in][]{Rezzolla_book:2013}. To obtain a set of conservation
equations suitable for numerical integration, and as is customary for
ordinary GRMHD, we can make use of a $3+1$ foliation of spacetime to
project the radiation energy momentum tensor onto (and in the direction
orthogonal to) a spacelike hypersurface, which yields
\begin{equation} \label{eq:tmunu_rad_euler}
  T^{\alpha \beta}_\text{rad} = E\, n^\alpha n^\beta + F^\alpha
  n^\beta + F^\beta n^\alpha + P^{\alpha\beta}\,.
\end{equation}
Here, $E$ is the radiation energy density and, by construction, $F_i$ is
a spatial vector and $P_{ij}$ is a symmetric rank-two tensor lying in the
tangent space to the spatial hypersurface. The equations for the grey M1
scheme finally read~\citep{Shibata2011}
\begin{align}
  \label{eq:M1_E}
\begin{split}
	\partial_t \left( \sqrt{\gamma} E \right) &+ \partial_j \left(
        \alpha \sqrt{\gamma} F^j - \sqrt{\gamma}\beta^j E \right) =
        \\ &\sqrt{\gamma} \left( \alpha P^{ij} K_{ij} - F^j\partial_j
        \alpha \right) - \alpha \sqrt{\gamma} S^\mu n_\mu\,,
\end{split}
\end{align}
\begin{align}
  \label{eq:M1_F}
\begin{split}
  \partial_t \left( \sqrt{\gamma} F_i \right) &+ \partial_j \left( \alpha
  \sqrt{\gamma} P^j_i - \sqrt{\gamma}\beta^j F_i \right) =
  \\ &\sqrt{\gamma} \left( -E\partial_i\alpha + F_j\partial_i\beta^j +
  \left(\frac{\alpha}{2}\right)\ P^{jk}\partial_i \gamma_{jk} \right)
  +\alpha \sqrt{\gamma} S^\mu \gamma_{\mu i}\,,
\end{split}
\end{align}
where, as customary, $\beta^i$ is the shift vector, $\alpha$ the lapse,
$\gamma_{ij}$ the spatial three-metric, and $\gamma$ its determinant.
We also define the Lorentz factor $W$ associated with the fluid
four-velocity $u^\alpha$ as $W:=\alpha u^t$.

Hence, in addition to the standard GRMHD equations in our
code~\citep{Most2019b}, we evolve equations \eqref{eq:M1_E} and
\eqref{eq:M1_F} for three effective neutrino species: the electron
neutrinos $\nu_e$, the electron anti-neutrinos $\bar{\nu}_e$, and an
effective heavy neutrino species $\nu_x$. While the approximation of
evolving the energy integrated moments (grey) is crucial to attain the
desired computational efficiency, it foregoes any information regarding
the neutrino energy spectrum, which might be extremely important to
correctly capture the composition of matter in the aftermath of a BNS
merger. Moreover, the set of Eqs.~\eqref{eq:M1_E}, \eqref{eq:M1_F} does
not guarantee the conservation of the total lepton number. To
partially overcome these issues and following~\citet{Foucart2016a,
  Radice2022}, we evolve an additional equation for the neutrino number
density. In particular, we introduce an effective neutrino number current
$N^\alpha$ and assume that the flow of neutrino number comoves with the
neutrino energy flux, \ie
\begin{equation}
  \label{eq:M1_n_def}
  N^\alpha = \tilde{n} \left( u^\alpha + \frac{H^\alpha}{J} \right)\,,
\end{equation}
where we have defined the neutrino number density in the fluid frame as
$\tilde{n}$. Defining then
\begin{equation}
  \label{eq:Gamma_N}
  \Gamma := \frac{n_\mu N^\mu}{\tilde{n}} = W \frac{E - F^i v_i}{ J }\,,
\end{equation}
with the last equality following from the fact that $H^\mu$ is orthogonal
to $u^\mu$ ($H^{\mu}u_{\mu}=0)$, we can write the evolution equation for
the neutrino number density as
\begin{equation}
  \label{eq:M1_N}
  \partial_t\left( \sqrt{\gamma}\, \Gamma \tilde{n} \right) + \partial_j
  \sqrt{\gamma} \alpha \tilde{n} \left( W\left( v^j - \frac{\beta^j}{\alpha}
  \right) + \frac{H^j}{J} \right) = \alpha\sqrt{\gamma} \mathcal{N}\,.
\end{equation}
where the source term $\mathcal{N}$ will be presented below
[see Eq.~\eqref{eq:source_N}] and $v^i$ is the projection of $u^\alpha$
in the tangent space of the three-hypersurface
\begin{equation} \label{eq:v_def}
  v^i = \frac{1}{W}\,\gamma^i_\alpha u^\alpha = \frac{1}{\alpha}\left( \frac{u^i}{u^0} + \beta^i \right) \,.
\end{equation}
Note also that in writing Eq.~\eqref{eq:M1_N} we have made the
implicit assumption that the neutrinos have the same average energy
in the advection term as in the energy-flux term.
This is a valid assumption if all the neutrinos have
the same energy and it is not strictly true in general; however, it
represents a reasonable approximation in most of the cases we are
interested in.

We now turn our attention to the source terms, which provide the coupling
between the radiation and the ordinary fluid. These terms will model
phenomena of isotropic, elastic scattering of neutrinos with the fluid,
absorption and the emission of radiation. In particular, we can write the
source terms as
\begin{equation} \label{eq:sources_vector}
  S^\alpha = (Q^{\rm er} - \kappa^{\rm er}_a J)\,u^\alpha - (\kappa^{\rm er}_a + \kappa^{\rm er}_s)
  \, H^\alpha\,,
\end{equation}
so that, in the Eulerian frame, this yields
\begin{equation}
  \label{eq:sources_eul}
  \boldsymbol{\bar{G}}^\alpha := \begin{pmatrix} -S^\mu n_\mu \\ ~ \\ S^\mu
    \gamma_{\mu i} \end{pmatrix} = \begin{pmatrix}  W \left( Q^{\rm er} +
    \kappa^{\rm er}_s J - \kappa^{\rm er} (E - F^i v_i) \right) \\ ~ \\ \left( W(Q^{\rm er} -
    \kappa^{\rm er}_a)v_i - \kappa^{\rm er} \gamma_{\mu i} H^\mu \right)\end{pmatrix}\,,
\end{equation}
which will serve as coupling to the energy and momentum conservation
equations. The RHS of Eq.~\eqref{eq:M1_N}, which will be coupled to the
fluid's electron fraction, reads
\begin{equation}
  \label{eq:source_N}
	\mathcal{N} = Q^{\rm nr} - \kappa^{\rm nr}_a \tilde{n}\,.
\end{equation}
In Eqs.~\eqref{eq:sources_eul},~\eqref{eq:source_N} above,
$Q^{\rm er}\,(Q^{\rm nr})$ refer to the total energy (number) emissivity
of a given neutrino species, while $\kappa^{\rm er}_a\,(\kappa^{\rm
  nr}_a)$ are the opacities due to energy (number) absorption and
$\kappa^{\rm er}_z\,(\kappa^{\rm nr}_z)$ the opacities due to scattering;
we additionally indicate the total opacity by
\begin{equation}
  \kappa^{\rm nr,er} :=
  \kappa^{\rm nr,er}_a+\kappa^{\rm nr,er}_s\,.
\end{equation}
In the following sections we will often forego the superscript and from
the context it will be understood whether we are referring to energy or
the number rates.

Note that, as written, Eqs.~\eqref{eq:M1_E} and \eqref{eq:M1_F} are
exact, but the corresponding system cannot be solved because it lacks an
evolution equation of the rank-two tensor $P^{ij}$. This is not unusual
in a moment-based scheme where the the solution of the lower-order
moments requires the knowledge of higher-order ones and the system is
complete only with an infinite expansion. The standard solution to this
otherwise unavoidable problem is via the introduction of a ``closure
relation'', that is, an equation relating higher-order moments with the
lower ones. In our case this amounts to prescribing a mathematically well
behaved and physically well-motivated relation for the pressure tensor of
the type
\begin{equation}
  P^{ij} = P^{ij}(E, F_i)\,.
\end{equation}
In deciding for such a closure we are aided by a large literature that
has experimented on prescriptions that allow us to smoothly express the
pressure tensor in the limit of an optically-thick medium, where the
radiation is in thermodynamic equilibrium with the fluid, and in the
optically-thin limit for radiation emitted by a point-like source. We
discuss below the details of the closures implemented in \FMONEB.

\subsection{Closure Relations}

As customary in implementations of the M1 scheme, we here adopt the following
ansatz for computing the neutrino pressure tensor which is expressed as a
linear weighted average between the expression it assumes in the optically-thin and
optically-thick regimes, namely
\begin{equation}
  \label{eq:closure}
  P^{ij} = d_\text{th} P^{ij}_\text{th} + d_{^{\text{TH}}} P^{ij}_{^{\text{TH}}}\,,
\end{equation}
where
\begin{align*}
	d_\text{th} &= \frac{3\chi - 1}{2}\,, \\
	d_{^{\text{TH}}} &= \frac{3(1-\chi)}{2} = 1-d_\text{th}\,.
\end{align*}
\FMONEA employs Minerbo's~\citep{Minerbo1978} closure
by default, for which
\begin{equation}
  \label{eq:minerbo}
  \chi := \frac{1}{3} + \xi^2\left( \frac{6 - 2\xi +6\xi^2}{15} \right)\,,
\end{equation} 
with $\xi$ being the variable Eddington factor defined as
\begin{equation}
  \label{eq:eddington_factor}
  \xi := \sqrt{ \frac{H^\mu H_\mu}{ J^2 } }\,,
\end{equation}
which is zero in the optically-thick limit and one in the optically-thin
one. Note that $\chi(\xi=0)=1/3$ and $\chi(\xi=1)=1$ and that
Eq.~\eqref{eq:eddington_factor} requires the knowledge of the fluid frame
moments of the radiation field since $H_i$ and $J$ appear in
Eq.~\eqref{eq:eddington_factor} as opposed to $F_i$ and $E$. This is due
to the fact that $F_i$ is not guaranteed to be small in the
optically-thick limit \citep{Shibata2011}. 

Because physical quantities related to the ordinary fluid or to the
radiation fluid are meaningfully expressed in the locally comoving frame,
but the actual evolution equations are expressed in terms of an Eulerian
frame, it is important to express the Eulerian-frame moments (we recall
that these are the radiation energy density $E$ and its flux
$F_{\alpha}$) in terms of the fluid-frame ones
\begin{align}
  \label{eq:E_fluid_gen}
    E := T^{\mu\nu} n_\mu n_\nu = & W^2 J - 2 W H^\mu n_\mu + L^{\mu\nu}
    n_\mu n_\nu\,, \\
  \label{eq:F_fluid_gen}
    F_\alpha := -T^{\mu\nu} n_\mu \gamma_{\nu\alpha} = & W^2 J v_\alpha +
    W H^\mu n_\mu \left( n_\alpha - v_\alpha \right) +
    \nonumber\\ &\phantom{-} WH_\alpha + L^{\mu\nu} n_\mu \gamma_{\nu
      \alpha}\,.
\end{align}
Using again the fact that $H^\alpha$ is orthogonal to $u^\alpha$, we can
contract Eq.~\eqref{eq:F_fluid_gen} with $\boldsymbol{u}$ and find
\begin{align}
  \label{eq:Fu_fluid_gen}
    F_\mu u^\mu = &WJ\left( W^2 -1 \right) + H^\mu n_\mu ( 1 - 2W^2) -
    L^{\mu\nu} n_\mu ( u_\nu - W n_\nu) \nonumber\\ = & W E + H^\mu n_\mu - WJ\,,
\end{align}
from which we can obtain an expression for the projection of $H^\alpha$
along the unit normal $\boldsymbol{n}$
\begin{equation}
  \label{eq:Hn}
  H^\mu n_\mu = W \left( F^i v_i -  E + J \right)\,.
\end{equation}
Finally, we can compute the pressure tensor as
\begin{align}
  \label{eq:P_fluid_gen}
  \begin{split}
    P_{\alpha\beta} := \gamma_{\mu\alpha}\gamma_{\nu\beta} T^{\mu\nu} = &
    W^2Jv_\alpha v_\beta + W \gamma_{\mu\alpha}H^\mu v_\beta + \\ &
    W\gamma_{\nu\beta} H^\nu v_\alpha + \gamma_{\mu\alpha}\gamma_{\nu\beta}
    L^{\mu\nu}\,.
  \end{split}
\end{align}

Specialising now to the optically-thick case, for which we can assume the
pressure tensor to be isotropic in the fluid rest-frame
\begin{equation}
  \label{eq:L_thick}
  L^{\alpha \beta} = \frac{1}{3} J h^{\alpha\beta} = \frac{1}{3} J \left(
  g^{\alpha \beta} + u^\alpha u^\beta \right)\,,
\end{equation}
we then use Eqs.~\eqref{eq:E_fluid_gen} and \eqref{eq:F_fluid_gen} to
write 
\begin{align} 
  E = &\frac{1}{3} J \left(4W^2 -1 \right) - 2 W H^\mu
  n_\mu \label{eq:E_Eulerian_thick}\,, \\
  \label{eq:F_Eulerian_thick}
    F_\alpha = &\frac{4}{3}W^2 J
    v_\alpha + WH^\mu n_\mu ( n_\alpha - v_\alpha ) + 
    WH_\alpha\,,
\end{align}
and Eq.~\eqref{eq:Hn} to express $H^\mu n_\mu$ in Eq.~\eqref{eq:E_Eulerian_thick}
to obtain
\begin{equation}
  \label{eq:J_thick}
  J = \frac{3}{2W^2+1}  \left( (2W^2-1)E - 2W^2 \,F^i\,v_i \right)\,.
\end{equation}
We then employ Eqs.~\eqref{eq:F_Eulerian_thick} and \eqref{eq:J_thick} to
find
\begin{equation}
  \label{eq:H_thick}
  \gamma^\alpha_j H_\alpha = W^{-1} F_j + \frac{Wv_j}{2W^2 +
    1}\left[(4W^2+1)\,F^i\,v_i - 4W^2 E \right]\,,
\end{equation}
from which, using Eqs.~\eqref{eq:P_fluid_gen} and \eqref{eq:L_thick}, we
finally obtain the desired expression for the radiation-pressure tensor
in the optically-thick limit
\begin{equation}
  \label{eq:P_thick}
	P^{ij}_{^{\text{TH}}} = \frac{4}{3} W^2 J v^i v^j + W \gamma_\mu^i
        H^\mu v^j + W \gamma_\mu^j H^\mu v^i + \frac{1}{3} J
        \gamma^{ij}\,.
\end{equation}

Similarly, when dealing with the optically-thin case, we can express the
pressure tensor in the Eulerian frame for a point-like emitter as
\begin{equation}
  \label{eq:P_thin}
  P_\text{th}^{ij} = E \frac{F^iF^j}{{F_k} {F^k}} = E
  \hat{f}^i\hat{f}^j\,,
\end{equation}
where we introduced the normalised energy flux
\begin{equation}
\hat{f}_i:=\frac{F_i}{\sqrt{F_k F^k}}\,.
\end{equation}
Using expression~\eqref{eq:P_thin}, we can easily compute the fluid-frame
moments 
\begin{align}
  \label{eq:J_thin}
    J = T^{\mu\nu} u_\mu u_\nu = W^2 \left\{ \left[ 1 + {(\hat{f}^i \, v_i )}^2 \right]\, E - 2 (F^i\,v_i) \right\} \,,
\end{align}
and
\begin{align} \label{eq:H_thin}
  \begin{split}
    \gamma^\alpha_j H_\alpha = \gamma^\alpha_j T^{\mu\nu} u_\mu
    h_{\nu\alpha} = & W^3 \left[ 2\,F^i\,v_i - E \right] v_j + W F_j \\ &
    - W E {( \hat{f}^i \, v_i }) \hat{f}_j - W^3 E {\left( \hat{f}^i\, v_i
      \right)}^2 v_j\,.
  \end{split}
\end{align}

We are now in a position to write down the fluid frame moments in terms
of the Eulerian frame ones. From Eqs.~\eqref{eq:J_thick},
\eqref{eq:J_thin} we then obtain 
\begin{equation} \label{eq:J_gen}
  J = B_0 + d_\text{th} B_\text{th} + d_{^{\text{TH}}} B_{^{\text{TH}}}\,, 
\end{equation}
and from Eqs.~\eqref{eq:H_thick}, \eqref{eq:H_thin}, and \eqref{eq:Hn}
\begin{align} \label{eq:H_gen}
  \begin{split}
    H^\alpha =& ( b_n^0 + d_\text{th} b_n^\text{th} + d_{^{\text{TH}}} b_n^{_{\text{TH}}} ) n^\alpha 
    -( b_v^0 + d_\text{th} b_v^\text{th} + d_{^{\text{TH}}} b_v^{_{\text{TH}}} ) v^\alpha \\ 
    & - d_\text{th} b_f^\text{th} \hat{f}^\alpha - ( b_F^0 + d_{^{\text{TH}}} b_F^{_{\text{TH}}} ) F^\alpha\,,
  \end{split}
\end{align}
where 
\begin{align}
  \label{eq:J_coeffs}
  &B_0 := W^2\left( E - 2 F^i v_i \right)\,,\\
  &B_\text{th} := W^2 E {\left( \hat{f}^i \,v_i \right)}^2\,,\\
  &B_{^{\text{TH}}} := \frac{W^2-1}{2W^2 + 1 }\left[ 4W^2 F^i v_i + \left(3 - 2W^2\right)E \right]\,,
\end{align}
and
\begin{align}
  \label{eq:H_coeffs}
  &b_n^0 := WB_0 + W\left( F^i v_i - E  \right)\,,\\
  &b_n^\text{th} := W B_\text{th}\,,\\
  &b_n^{_{\text{TH}}} := W B_{^{\text{TH}}}\,,\\ 
  &b_v^0 := b_n^0 \\
  &b_v^\text{th} := b_n^\text{th}\,,\\
  &b_v^{_{\text{TH}}} := W B_{^{\text{TH}}} + \frac{W}{2W^2+1}\left[ (3-2W^2)E + (2W^2-1)F^i v_i \right]\,,\\ 
  &b_F^0 := -W\,,\\
  &b_f^\text{th} := W E \left( \hat{f}^i\,v_i \right) \,,\\
  &b_F^{_{\text{TH}}} := W v^i v_i\,. 
\end{align}

We now have all the information needed to compute the closure. As can be
seen from Eqs.~\eqref{eq:J_gen} and \eqref{eq:H_gen}, the computation of
$\xi$ is a nonlinear problem, where the knowledge of the evolved moments
$E, F_\alpha$ is not sufficient to find the closure and a numerical
root-finding approach is necessary. Following \citet{Foucart2015a,
  Weih2020b}, in \FMONEA this is done by looking for the root of
\begin{equation}
  \label{eq:closure_rootfinding_func}
	R(E, F_i) = \frac{H^\mu H_\mu - J^2}{E^2}\,,
\end{equation}
via a Newton-Raphson method in conjunction with a Brent method as a
fallback in case the first approach fails.

It is worth pointing out here that the choice of a closure function in
the form of a relation $\chi = \chi(\xi) \in [1/3,1]$ is not unique and
indeed it represents one of the main sources of uncertainty in the M1
scheme and more in general in any truncated formalism approach
\citep{Rezzolla1994}. The choice of the Minerbo closure, which is the
classical form of the maximum-entropy closure, is made
following~\citet{Murchikova2017}, where it was identified as the closure
which performed best on average in a series of tests among a series of
possible analytic closures. \FMONEA, however, leaves the option of
implementing additional closures and comes equipped with the the
Levermore closure~\citep{Levermore1984}. 

\section{Numerical scheme}
\label{sec:numerics}

In this section we provide a detailed description of the numerical
algorithm employed by \FMONEB. As a first step, it is worth remarking
that the full set of the GRMHD and radiative-transfer equations
represents a system of nonlinear partial differential equations that can
be cast in a conservative form~\citep{Rezzolla_book:2013}
\begin{equation}
  \label{eq:full_conservative}
  \partial_t
  \boldsymbol{{U}} + \partial_j
  \boldsymbol{{F}}^j(\boldsymbol{{U}}) =
  \boldsymbol{{S}}(\boldsymbol{{U}}) \,,
\end{equation}
where $\boldsymbol{{U}}$ is the state vector and contains all of the
conserved quantities that are evolved in time, while $\boldsymbol{{F}}^j$
and $\boldsymbol{S}$ are the fluxes and source terms, respectively. For
compactness, we will concentrate here only on the subset of variables
that are relevant for the radiative-transfer portion of the system and
remind the interested reader to the various papers where the
corresponding GRMHD part is presented~\citep[see, \eg][]{Fambri2018}. In
this case, the system of equations can still be cast in the conservative
form~\eqref{eq:full_conservative}, but the state vector will be just a
part of the full state vector and we will distinguish the source term
$\boldsymbol{S}$ into a part that is related to the radiation quantities,
and that we indicate with $\boldsymbol{\tilde{G}}$, and another one that
contains instead also information on the spacetime metric and extrinsic
curvature, that we refer to as $\boldsymbol{\tilde{S}}$. As a result, the
conservative form of the radiative-transfer equations reads
\begin{equation}
  \label{eq:M1_conservative}
  \partial_t
  \boldsymbol{\tilde{U}} + \partial_j
  \boldsymbol{\tilde{F}}^j(\boldsymbol{\tilde{U}}) =
  \boldsymbol{\tilde{S}}(\boldsymbol{\tilde{U}}) +
  \boldsymbol{\tilde{G}}(\boldsymbol{\tilde{U}}) \,,
\end{equation}
where
\begin{equation}
  \label{eq:M1_U}
  \boldsymbol{\tilde{U}} := \begin{pmatrix} \sqrt{\gamma} \Gamma \tilde{n}
          \\ ~ \\ \sqrt{\gamma} E \\ ~ \\ \sqrt{\gamma} F_i \end{pmatrix}\,,
\end{equation}
and [see Eq.~\eqref{eq:sources_eul} for a definition of
  $\boldsymbol{\bar{G}}^{\alpha}$]
\begin{equation}
  \label{eq:M1_G}
  \boldsymbol{\tilde{G}} := \sqrt{\gamma} \alpha\, \begin{pmatrix}\mathcal{N}
    \\ ~ \\ \boldsymbol{\bar{G}}^\alpha \end{pmatrix}\,,
\end{equation} 
with
\begin{equation}
  \label{eq:M1_F_1}
  \boldsymbol{\tilde{F}}^j := 
  \sqrt{\gamma} \alpha \,\begin{pmatrix}
     \tilde{n} \left( W\left( v^j 
    - {\beta^j}/{\alpha} \right) + {H^j}/{J} \right) \\ ~ \\
    F^j - {\beta^j}E/{\alpha}  \\ ~ \\
    P^j_i - {\beta^j}F_i/{\alpha} 
  \end{pmatrix}\,,
\end{equation} 
and
\begin{equation}
  \label{eq:M1_S}
  \boldsymbol{\tilde{S}} := 
  \sqrt{\gamma} \,\begin{pmatrix}
     0 \\ ~ \\
     \alpha P^{ij} K_{ij} - F^i\partial_i \alpha \\ ~ \\
     -E\partial_i\alpha + F_j\partial_i\beta^j +
     {\alpha}\ P^{jk}\partial_i \gamma_{jk}/{2}
  \end{pmatrix}\,.
\end{equation}

The importance of casting the system in the conservative
formulation~\eqref{eq:M1_U} is that can use standard high-resolution
shock-capturing (HRSC) methods to numerically solve
Eqs.~\eqref{eq:M1_conservative}. In particular: \FMONEA discretizes the
fluxes in Eqs.~\eqref{eq:M1_conservative} by standard second-order
accurate finite-volume techniques. Special care has to be taken when
dealing with the M1 system for two reasons:
\begin{enumerate}
\item when the mean free path of the particles tends to zero, the
  equations tend asymptotically to diffusion equations for the energy and
  number density of neutrinos. Standard techniques for hyperbolic
  equations do not reproduce this limit accurately introducing instead
  significant amounts of numerical dissipation in the solution;
\item 
  in hot and dense regions as those normally occurring in BNS
  simulations, the collisional source terms $\boldsymbol{\tilde{G}}$ can
  become stiff leading to inaccurate solutions.
\end{enumerate}
We discuss below how to address these issues in practice.

\subsection{Flux discretization}

As anticipated, \FMONEA discretizes the equations via second-order
accurate, HRSC finite-volume methods. Given a conserved variable $u^n_i$
discretized at time $t^n$ and at a spatial point $x_i$, we write the
semidiscrete equivalent of the system~\eqref{eq:M1_conservative} as
\begin{equation*}
\partial_t \tilde{\mathcal{U}}^n_i = -\frac{1}{h} \left( \mathcal{F}^{j}_{i + 1/2} -
\mathcal{F}^{j}_{i - 1/2} \right) + \mathcal{S}(\mathcal{U}^n_i) + \dots\,,
\end{equation*}
where the dots indicate the collisional sources whose treatment will be
detailed in sec.~\ref{ssec:collisional}. The numerical fluxes
$\mathcal{F}^j$ are computed at cell interfaces based on the
reconstructed values of the primitive variables for the M1 scheme. In
\FMONEA we reconstruct $\left( N/E, E, F_i/E \right)$ to ensure the
causality of the energy fluxes using the standard second-order
total-variation diminishing (TVD) monotonized central reconstruction
scheme~\citep{Foucart2015a}. The code has the option of also using a
simple MinMod reconstruction or a third-order accurate WENO
reconstruction method~\citep[see, \eg][for
  details]{Rezzolla_book:2013}. To compute the fluxes, we utilise the
two-wave Harten-Lax-van Leer-Einfeldt (HLLE) approximate Riemann
solver~\citep{Harten83, Einfeldt88} with eigenspeeds~\citep{Foucart2015a,
  Shibata2011, Weih2020b} which read
\begin{equation}
  \label{eq:m1_speeds}
  \lambda^{(\pm)} := d_\text{th} \lambda^{(\pm)}_\text{th} +
  d_{^{\text{TH}}} \lambda^{(\pm)}_{^{\text{TH}}}\,,
\end{equation}
where
\begin{align}
  \label{eq:lambda_thick}
  & \lambda^{(\pm)}_{^{\text{TH}}} := \min\left(-\beta^j + p^j, r^j \right)\,, \\
  & p^j := \frac{\alpha v^j}{W}\,, \\
  & r^j := -\beta^j + \frac{2W^2p^j \pm \sqrt{\alpha^2\gamma^{jj}\left( 2W^2+1\right)-2\left( Wp^j\right)^2} }{2W^2+1}\,,
\end{align}
and
\begin{equation}
  \label{eq:lambda_thin}
  \lambda^{(\pm)}_\text{th} := -\beta^j \pm \alpha
         {\frac{F^j}{\sqrt{F^i F_i}}}\,.
\end{equation}

As mentioned above, when computing the numerical fluxes at cell
interfaces we need to ensure that the scheme is asymptotically
preserving: \ie we need to correct the fluxes in a way that will make
them suitable to solve a diffusion-type equation in the optically-thick
limit. Several approaches to this problem have been proposed in the
literature. One possibility, employed for instance
by~\citet{Foucart2015a} and \citet{Weih2020b}, is to interpolate between
the HLLE fluxes in the optically-thin regime and a diffusive flux
obtained from a finite-difference approximation of the derivative of $J$
in the optically-thick one. This approach, while shown to yield good
results, is cumbersome and requires the numerical differentiation of an
expression which itself contains a derivative. On the other
hand,~\citet{Radice2022} correct the fluxes by using centered differences
in the optically-thick regime and employing a flux limiter to hybridise
the high-order centered flux with a diffusive flux near shocks or extrema
of the solution. While this second approach is computationally less
expensive, it neglects altogether the causal information coming from the
eigenvalues of the system. For these reasons, in \FMONEA we apply the
following correction to the HLLE fluxes for energy and number
densities~\citep[see also][for a similar
  approach]{Kuroda2016,Skinner2019,Cheong2023}
\begin{equation}
  \label{eq:HLLE_E_mod}
  \mathcal{F}^\text{HLLE-mod}_{i+1/2} = \frac{ \lambda^{(+)} F_L -
    \lambda^{(-)} F_R + A \lambda^{(+)} \lambda^{(-)} (U_R + U_L) }{
    \lambda^{(+)} + \lambda^{(-)} }\,,
\end{equation}
where the diffusion-limiter factor $A$ is defined as 
\begin{equation}
  \label{eq:A_def}
  A := \min\left( 1, \frac{1}{\Delta x \langle\kappa\rangle} \right)\,,
\end{equation}
and represents effectively is the mean-free-path of neutrinos computed
with an opacity $\langle\kappa\rangle$ which is taken to be the average
among the two neighbouring cells at the interface. For the energy fluxes $F_i$
we instead apply the simpler correction given by
\begin{equation}
  \label{eq:F_flux_mod}
  \mathcal{F}^j_{i + 1/2} = \frac{1}{2}\left( A^2
  \mathcal{F}^{\text{HLLE},j}_{i + 1/2} + \frac{1}{2}\left(1-A^2\right) \left( F^j_R
  + F^j_L \right)\right)\,.
\end{equation}
Expression~\eqref{eq:F_flux_mod} effectively switches off the
diffusive term in the HLLE solver at high opacities, turning the fluxes
into a form similar to a centered-differencing scheme (where however the
fluxes are computed at the left and right side of the interface as opposed to the cell centres of neighbouring cells).

\subsection{Collisional sources}
\label{ssec:collisional}

As anticipated above, the reaction rates (emissivity, opacities) involved
in the computation of the collisional sources, and which serve as
coupling between the radiation and the fluid, can become very large in
the hot and dense matter found in BNS mergers~\citep{Endrizzi2020}. As a
result, the collisional sources in Eq.~\eqref{eq:M1_conservative} are not
only a nonlinear function of the evolved variables but can also become
very stiff. For this reason it is common for M1 schemes \citep{Weih2020b,
  MelonFuksman2019, Foucart2015a, McKinney2014} to employ
Implicit-Explicit Runge-Kutta time-integration schemes or
RK-IMEX~\cite{Pareschi2010, Palenzuela:2008sf}. In \FMONEA we
follow~\citet{Radice2022} and evolve the system in time according to
\begin{align}
  \label{eq:time_stepping_scheme}
    \boldsymbol{\tilde{U}}^{(*)} = &\boldsymbol{\tilde{U}}^{(n)} + \Delta t \left[
      -\partial_i \boldsymbol{\tilde{F}}\left(\boldsymbol{\tilde{U}}^{(n)}\right) +
      \boldsymbol{\tilde{S}}\left(\boldsymbol{\tilde{U}}^{(n)}\right) +
      \boldsymbol{\tilde{G}}\left(\boldsymbol{\tilde{U}}^{(*)}\right) \right]\,,
    \\ \boldsymbol{\tilde{U}}^{(n+1)} = &\boldsymbol{\tilde{U}}^{(*)} + \Delta t \left[
      -\partial_i \boldsymbol{\tilde{F}}\left(\boldsymbol{\tilde{U}}^{(*)}\right) +
      \boldsymbol{\tilde{S}}\left(\boldsymbol{\tilde{U}}^{(*)}\right) +
      \boldsymbol{\tilde{G}}\left(\boldsymbol{\tilde{U}}^{(n+1)}\right) \right]\,,
\end{align}
where the the upper index $(*)$ is employed to indicate an intermediate
step in the time evolution and where fluid is updated at the end of the
second substep. We then need to solve the nonlinear system
\begin{equation}
  \label{eq:implicit_eq}
  \boldsymbol{\tilde{X}}\left(\boldsymbol{\tilde{U}}^{(n)}\right)+
  \boldsymbol{\tilde{G}}\left(\boldsymbol{\tilde{U}}^{(*)}\right)-\boldsymbol{\tilde{U}}^{(*)}=0\,,
\end{equation}
where we indicate by $\boldsymbol{\tilde{X}}$ the vector of partially
updated variables (\ie where the flux and geometric source terms have
already been added). Since the reaction rates are kept fixed through the
neutrino update, the source term for the neutrino number density
decouples from the rest of the system and can be inverted
analytically. Following~\citet{Radice2022}, we solve this by a globally
convergent Newton-Raphson procedure where the full Jacobian of the system
is evaluated. In particular, the equation for the collisional source term
of the neutrino number density, \ie Eq.~\eqref{eq:source_N}, is linear in
the evolved variable and its inversion is trivial. We therefore only need
to focus on the subsystem that comprises the sources for $E$ and $F_i$,
and Eq.~\eqref{eq:sources_eul}, that is, a system of four coupled
nonlinear equations in the evolved radiation moments. The Jacobian can
be formally written down as
\begin{equation}
  \label{eq:jac_0}
  \boldsymbol{J} = \frac{d \boldsymbol{\bar{G}}}{d \boldsymbol{\bar{U}} } , 
\end{equation}
with $\boldsymbol{\bar{U}}$ being defined as the sub-vector of evolved
radiation variables where the number density has been excluded.  We thus
have
\begin{align}
  \boldsymbol{J}_0^0 &:= \frac{d \boldsymbol{\bar{G}}_0 }{d E}   =
  -\alpha W \left( \kappa^{\rm er} - \kappa_s^{\rm er} \frac{\partial J}{\partial E} \right)\,, \label{eq:J00} \\
  \boldsymbol{J}_0^j &:= \frac{d \boldsymbol{\bar{G}}_0 }{d F_j} =
  \alpha W \left( \kappa_s^{\rm er} \frac{\partial J}{\partial F_j} + \kappa^{\rm er}\, v^j \right)\,, \label{eq:J0j} \\
  \boldsymbol{J}_k^0 &:= \frac{d \boldsymbol{\bar{G}}_k }{d E} =
  -\alpha \left( \kappa^{\rm er} \frac{\partial H_k }{ \partial E  } + W \kappa_a^{\rm er} \frac{\partial J}{\partial E} v_k \right)\,, \label{eq:Ji0} \\
  \boldsymbol{J}_k^j &:= \frac{d \boldsymbol{\bar{G}}_k }{d F_j} =
  -\alpha \left( \kappa^{\rm er} \frac{\partial H_k }{ \partial F_j  } + W \kappa_a^{\rm er} \frac{\partial J}{\partial F_j} v_k \right)\,. \label{eq:Jij} 
\end{align}
Note that the derivatives can be computed from Eqs.~\eqref{eq:J_gen}
and~\eqref{eq:H_gen}, and that the term in $H^\alpha$ proportional to the
hypersurface normal $\boldsymbol{n}$ does not contribute to the Jacobian,
since only $\gamma_{\mu i}H^\mu$ enters the collisional source terms. The
relevant derivatives are \citep[see also][]{Radice2022,Izquierdo2022}
\begin{align}
  \frac{\partial J}{\partial E} =& W^2\,\left( 1 + d_\text{th} \, (\hat{f}_i\,v^i)^2 \right) + d_\text{TH} \frac{ ( 3-2W^2)(W^2-1)}{ 1 + 2W^2}\,, \label{eq:dJdE} \\
  \frac{\partial H_j}{\partial E} =& -W^3\,\left( 1 + d_\text{th}\,(\hat{f}_i\,v^i)^2 - d_\text{TH} \frac{2W^2 - 3}{1 + 2W^2} \right)\, v_j \nonumber \\
  &- d_\text{th} W (\hat{f}_i\,v^i) \hat{f}_j \,, \label{eq:dHdE} \\
  \frac{\partial J}{\partial F_j} =& 2W^2\,\left( - 1 + d_\text{th}\,\frac{E (\hat{f}_i\,v^i)}{\sqrt{F_iF^i}} - 2 d_\text{TH} \frac{W^2 - 1}{1 + 2W^2} \right)\, v^j \nonumber \\
  &- 2 d_\text{th}\frac{W^2E\,(\hat{f}_i\,v^i)}{\sqrt{F_iF^i}}  \hat{f}^j \,, \label{eq:dJdF} \\
  \frac{\partial H_k}{\partial F_j} =& W\,\left( 1-d_\text{th}\frac{E (\hat{f}_i\,v^i)}{\sqrt{F_iF^i}} - d_\text{TH} \,v_iv^i \right) \delta_k^j \nonumber \\
  &+ 2W^3\left[ 1 - d_\text{th}\,\frac{E (\hat{f}_i\,v^i)}{\sqrt{F_iF^i}} 
    - d_\text{TH}\left( 1 - \frac{1}{2W^2(1+2W^2)} \right) \right] v_k v^j \nonumber \\
  &+ 2W d_\text{th}\frac{E (\hat{f}_i\,v^i)}{\sqrt{F_iF^i}} \hat{f}_k\hat{f}^j + 2W^3 d_\text{th} \frac{E (\hat{f}_i\,v^i)}{\sqrt{F_iF^i}} v_k \hat{f}^j \nonumber \\
  &- W d_\text{th} \frac{E (\hat{f}_i\,v^i)}{\sqrt{F_iF^i}} \hat{f}_k v^j \,. \label{eq:dHdF} 
\end{align}
As an initial guess, we also use the first-order in $v$ approximation to
the sources, which in the fluid frame reads
\begin{align}
  \label{eq:solver_initguess}
  \begin{split}
    &\tilde{J} = \frac{J + Q{\Delta t}/{W}}  {1 + \kappa^{\rm er}_a{\Delta t}/{W} }\,, \\ 
    &\tilde{H}_i = \frac{ H_i } { 1 + ( \kappa^{\rm er}_a + \kappa^{\rm er}_s ){\Delta t}/{W} }\,,
  \end{split}
\end{align}
where $J$ and $H_i$ are evaluated after the explicit update. To transform
back to the evolved variables, the closure is assumed to be optically
thick~\citep{Radice2022}, Eqs.~\eqref{eq:E_Eulerian_thick} and
\eqref{eq:F_Eulerian_thick} are used and the projection
$\boldsymbol{H}\cdot \boldsymbol{n}$ can be found via the orthogonality
to $u$ and is simply given by
\begin{equation*}
	H^\mu n_\mu = - H^i v_i\,.
\end{equation*}  
In particular, the process of inverting the implicit sources in
\FMONEA proceeds as follows
\begin{enumerate}
\item The closure is updated using the Eulerian-frame moments already
  updated according to the explicit part of the scheme. At this point,
  the moments are also limited to ensure that the energy density is
  larger than a small but positive ``atmosphere'' value and the energy
  flux is limited to ensure that the flux factor (the ratio $|F|/E$ of the energy 
  flux to the energy density) is not greater than
  unity (which would otherwise break causality);
\item The solver is initialised according to the initial guess in
  Eq.~\eqref{eq:solver_initguess};
\item One iteration of the inversion method is performed; 
\item The closure is updated again and the moments checked for
  consistency with the conditions of point (i), namely that the energy
  density is above atmosphere and the fluxes respect causality; 
\item Upon convergence, the source term of the neutrino number density is
  computed based on the updated neutrino moments.
\end{enumerate}

\subsection{Transport Rates}
\label{sec:rates}

What remains to be fixed at this point are the transport rates dictating
the coupling of the radiation field to the fluid, \ie the
\textit{absorption opacity} $\kappa^{\rm er}_a$, the \textit{scattering opacity}
$\kappa^{\rm er}_s$, the \textit{emissivity} $Q$ and the corresponding number
rates. Since \FMONEA is a grey M1 scheme, the reaction rates need to be
appropriately averaged over the neutrino energies. In particular, from
the energy-integrated radiation transport equation
\begin{align}
  \kappa &:= \frac{ \int \kappa_{(f)} \mathscr{F} f^3 df}{ \int
    \mathscr{F} f^3  df }\,,\\
  Q &:= \int Q_{(f)}  \mathscr{F} f^3 df\,,
\end{align}
where $\mathscr{F}$ represents the radiation distribution function.

Table~\ref{tab:reactions} offers a list of all the reactions employed in
\FMONEA and the energy-integrated rates for all these processes can be
estimated analytically as a function of the local thermodynamic state of
the fluid (explicit expressions will be given in
Appendix~\ref{sec:appendix}). Furthermore, the chemical potential of
neutrinos as a function of $(\rho,T,Y_e)$ is calculated, as is customary
for radiative-transfer codes, assuming cold beta-equilibrium
\begin{align}
  \label{eq:nue_mu_eq}
  \mu^{\rm eq}_{\nu_e} &=  \mu_{e^-} + \mu_p - \mu_n - Q_{np}= - \mu^{\rm eq}_{\bar{\nu}_e}\,, \\
  \mu^{\rm eq}_{\nu_x} &= 0\,, \label{eq:nux_mu_eq}
\end{align}
where $Q_{np}$ indicates the nucleon mass difference and all the chemical
potentials $\mu_*$ include the rest mass contribution.
While this is appropriate for cold neutron-star matter, significant
corrections should be applied in high temperature regions of the EOS
table which in principle are probed during BNS merger simulations
\citep[see, \eg the discussion in][]{Hammond:2021vtv}.

Many leakage codes in the literature employ a corrected version of this
equilibrium chemical potential which is suppressed by a factor
$(1-\-\tau))$ where $\tau$ is the optical depth. This correction ensures
that the chemical potential goes to zero as matter becomes optically
thin. Analogously to \citep{Foucart2015a} we find that the choice
between $\mu^{\rm eq}$ and the ``leakage'' prescription has very little
impact on the simulations performed with M1, and we employ the corrected
chemical potential by default.
\begin{table}
  \centering
  \caption{List of weak reactions employed in \FMONEB. The index $i$
    stands for the different neutrino species, namely electron neutrino
    and anti-neutrino and effective heavy-lepton neutrino.}
  \label{tab:reactions}
  \begin{tabular}[width=.8\textwidth]{ll} 
    \hline
    \textbf{Scattering} & \\
    \hline
    Scattering on free nucleons & $\nu_i + N \rightarrow \nu_i + N,~ \bar{\nu}_i + N \rightarrow \bar{\nu}_i + N$ \\
    Scattering on heavy nuclei & $\nu_i + N \rightarrow \nu_i + N,~ \bar{\nu}_i + N \rightarrow \bar{\nu}_i + N$ \\
    \hline
    \textbf{Emission} & \\
    \hline
    Absorption on free nucleons & $\nu_e + n \rightarrow p + e^{-}, ~ \bar{\nu}_e + p \rightarrow n + e^{+}$ \\ 
    \hline
    \textbf{Emission} & \\
    \hline 
    Beta processes            & $p + e^- \rightarrow \nu_e + n ,~  n + e^+ \rightarrow \bar{\nu}_e+ p$ \\ 
    Transverse plasmon decay  & $\gamma \rightarrow \nu_x +\bar{\nu}_x$ \\ 
    Pair annihilation         & $e^- + e^+ \rightarrow \nu_x +\bar{\nu}_x$ \\ 
    Nucleon-nucleon           & $N+N \rightarrow \nu_x +\bar{\nu}_x$ \\  
    bremsstrahlung & \\
    \hline 
  \end{tabular}
\end{table}

Since the dominant charge current absorption rates are proportional to
the square average energy of neutrinos, we can employ the improved
knowledge about the neutrino spectrum coming from our effective evolution
of the neutrino number density to correct the free rates. Indeed, when
neutrinos decouple from the plasma and become free streaming, their
temperature no longer coincides with the local fluid temperature and
remains roughly constant. Because of this, computing the absorption and
scattering rates due to charge current reactions according to the local
thermodynamic state of the fluid leads to a large underestimation of the
neutrino transport rates in optically-thin regions. For this reason, and
in analogy with~\citet{Foucart2016a}, we apply the following correction
to the absorption opacities of electron-flavour neutrinos
\begin{equation}
   \label{eq:epscorr}
  \kappa^{\rm er,nr}_{a,~(\nu_e,\bar{\nu}_e)} =
  \left.\kappa^{\rm er,nr}_{a,~(\nu_e,\bar{\nu}_e)}\right|^{\rm eq} \ \max\left(1,
  \left(\frac{T_{(\nu_e,\bar{\nu}_e)}}{T}\right) \right)^2.
\end{equation}
An anlogous correction is applied to all the scattering opacities.
The neutrino temperature $T_\nu$ is defined as
\begin{equation}
  \label{eq:Tnu}
  T_\nu := \frac{\mathcal{F}_2(\eta_\nu)}{\mathcal{F}_3(\eta_\nu)} \, \langle \epsilon_\nu \rangle\,,
\end{equation}
where $\mathcal{F}_N(\eta)$ is the Fermi integral of order $N$ and is defined as
\begin{equation*}
  \mathcal{F}_N(\eta) := \int_{0}^{\infty} \frac{x^N}{e^{x-\eta} -1} dx\,,
\end{equation*}
and the average neutrino energy is extracted from the evolved variables
according to 
\begin{equation}
  \label{eq:epsnu}
	\langle \epsilon_\nu \rangle := W\frac{E - F_i v^i}{\tilde{n}}\,.
\end{equation}
After this temperature correction has been applied, and in order to
guarantee that the neutrinos are in thermal equilibrium with the fluid in
the optically trapped regions ($\tau\gg1$), we follow
\citep{Foucart2015a,Radice2022} and recompute the emission of
electron-flavour neutrinos according to an approximate form of Kirchhoff's
theorem given by
\begin{align}
  \label{eq:Kirchoff}
  &Q^{{\rm nr}} = \kappa^{{\rm nr}}_a\,B^{{\rm nr}}\left(T, \eta\right)\,, \\
  &Q^{{\rm er}} = \kappa^{{\rm er}}_a\,B^{{\rm er}}\left(T, \eta\right)\,, 
\end{align}
where $\eta$ is the neutrino fugacity and $B$ is the black-body spectrum
\begin{align}
&B^{\rm nr}_i\left(T, \eta\right) = g_i\,\frac{4\,\pi c}{(hc)^3} T^{3}
\mathcal{F}_{2}(\eta)\,, \\
&B^{\rm er}_i\left(T, \eta\right) = g_i\,\frac{4\,\pi c}{(hc)^3} T^{4} \mathcal{F}_{3}(\eta)\,,
\end{align}
with the statistical weight of the neutrino flavour $g_e=1,g_x=4$. In
particular, we set
\begin{align}
\label{eq:Q_ernr}
Q^{\rm er, nr}_{(\nu_e,\bar{\nu}_e)} &= B^{\rm er,nr}\left(T^*,\eta^*_{(\nu_e,\bar{\nu}_e)}\right) \kappa^{\rm er,nr}_{a,(\nu_e,\bar{\nu}_e)}\,,
\end{align}
and
\begin{align}
\label{eq:kappa_ernr}
\kappa^{\rm er,nr}_{a,(\nu_x)} &= \frac{Q^{\rm er, nr}_{(\nu_x)}}{B^{\rm
    er, nr}\left(T^*, \eta^*_{(\nu_x)}\right)}\,.
\end{align}

Note that the values of $T^*$ and $Y_e^*$
appearing in Eqs.~\eqref{eq:Q_ernr} and \eqref{eq:kappa_ernr} are set
depending on the timescale of beta equilibration. More precisely, we
estimate the beta-equilibration timescale as
\begin{equation*}
	\tau_{\beta-e} := \left[ \kappa^{\rm er}_a\,(\kappa^{\rm
          er}_a+\kappa^{\rm er}_s)\right]^{-1/2}\,,
\end{equation*}
and compare it with the simulation time step. If $\tau_{\beta-e}>\Delta
t$, then we simply set $(T^*,Y_e^*)=(T,Y_e)$. On the other hand, if
$\tau_{\beta-e}<\Delta t/2$, then the equilibration process cannot be
resolved by the simulation. Since the rates are updated only once per
timestep, and remain constant throughout the nonlinear solver iterations
used to determine the collisional source terms, using the local state of
the fluid can cause overshooting the equilibrium state and thus lead to
spurious oscillations that can be especially troublesome near black-hole
horizons or neutron-star surfaces. For this reason,
following~\citet{Radice2022}, we utilise instead of $(T,Y_e)$ their value
at beta equilibrium. Crucially, neutrino-less beta equilibrium is not
appropriate in this instance since the fluid is optically thick
($\tau_{\beta-e}\sim 1/\kappa$). We therefore need to perform a nonlinear
root-finding iteration to find $(T^*,Y_e^*)=(T^{\rm eq}, Y_e^{\rm
  eq})$. In particular, we solve the following
equations~\citep{Perego2019}
\begin{align} \label{eq:beta_eq_fix}
  Y_l &= Y^{\rm eq}_e + Y_{\nu_e}\left(T^{\rm eq}, Y_e^{\rm eq}\right) - Y_{\bar{\nu}_e}\left(T^{\rm eq}, Y_e^{\rm eq}\right)\,, \\ 
  u &= e\left(T^{\rm eq}, Y_e^{\rm eq}\right) + \frac{\rho}{m_b}\Big[ Z_{\nu_e}\left(T^{\rm eq}, Y_e^{\rm eq}\right)+ Z_{\bar{\nu}_e}\left(T^{\rm eq}, Y_e^{\rm eq}\right) \notag \\ 
    &\phantom{=}+ 4 \, Z_{\nu_x}\left(T^{\rm eq}, Y_e^{\rm eq}\right)\Big]\,,
\end{align}
where $e(T,Y_e) = (1+\rho)\epsilon(\rho,T,Y_e)$, with $\epsilon$ being
the specific internal energy computed from the EOS table, and $u, Y_l$
are defined as
\begin{align*}
	Y_l &:= Y_e + \frac{m_b}{\rho}\Big[ N_{\nu_e} - N_{\bar{\nu}_e} \Big]\,, \\
	u &:= e + E_{\nu_e} + E_{\bar{\nu}_e} + E_{\nu_x}\,.
\end{align*}
It is worth noting that this inversion requires interpolating the table
for $\epsilon$ at each iteration. Finally, in the intermediate the case
in which $1/2 < \tau_{\beta-e}/\Delta t < 1$, we determine $(T^*,Y_e^*)$
via a linear interpolation between the evolved and equilibrium values.

\begin{figure*}
\includegraphics[width=\columnwidth]{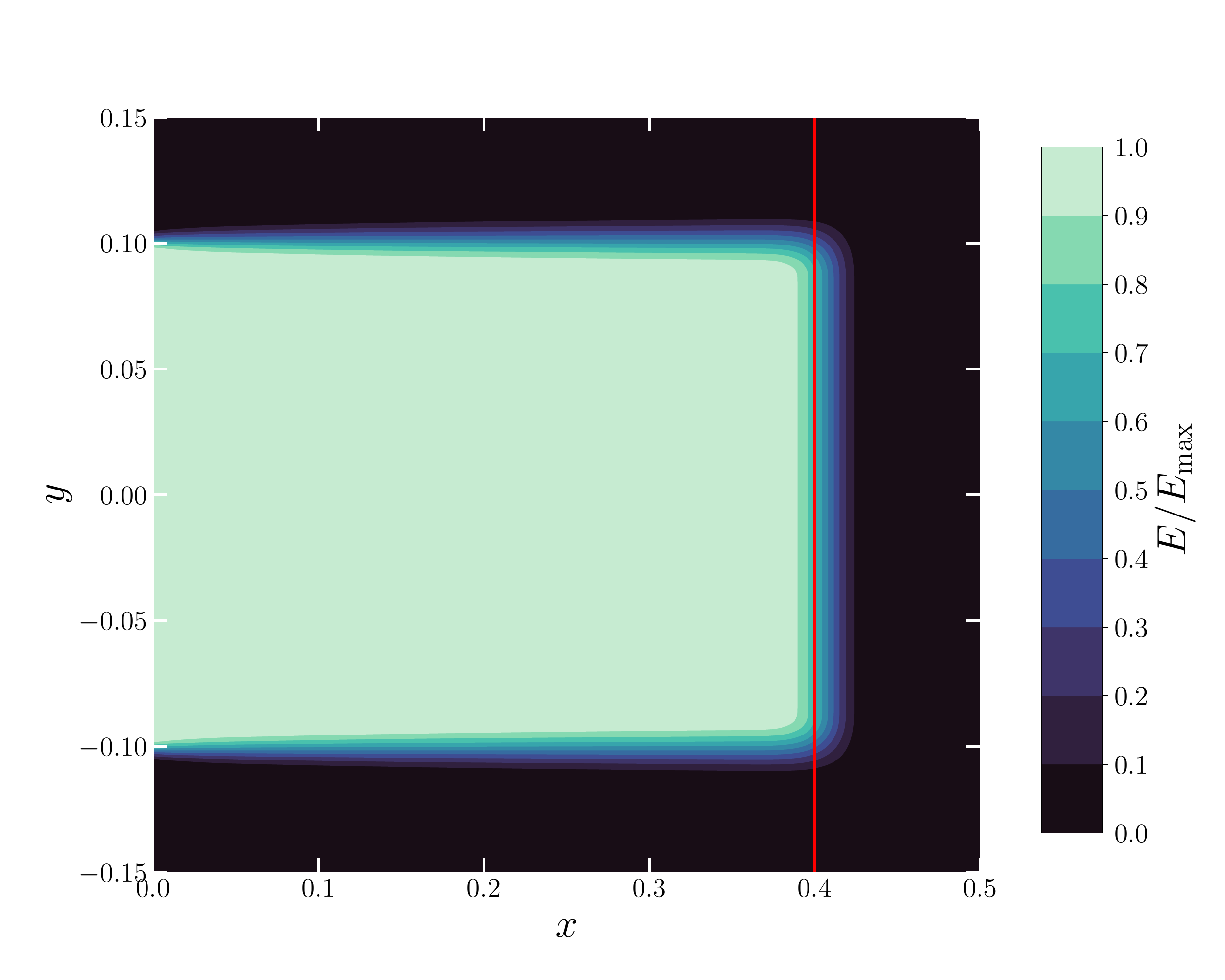}
\includegraphics[width=\columnwidth]{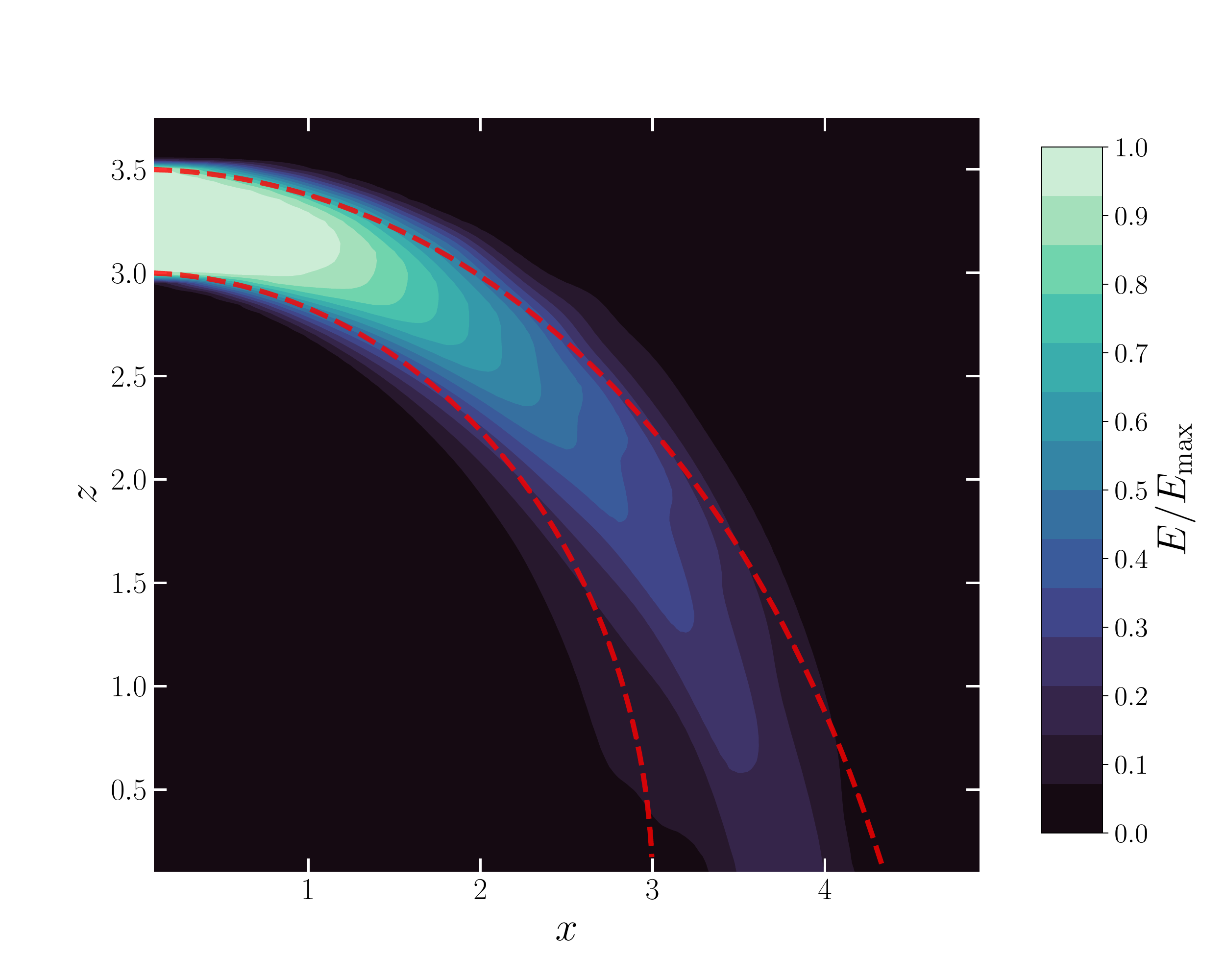}
\caption{\textit{Left:} Distribution at $t=0.4$ of the radiation energy
  density normalised to the maximum value for a radiation beam
  propagating in a Minkowski spacetime (straight-beam test). The expected
  position of the beam front is shown as a vertical red
  line. \textit{Right:} The same as in the left but for a radiation beam
  propagating near a Schwarzschild black hole. Red dashed lines indicate
  the exact geodesics at the edges of the injected beam.}
\label{fig:straight_beam}
\end{figure*}

\section{Implementation tests}
\label{sec:implementation_tests}

\FMONEA is written to be used in conjunction with the
\texttt{EinsteinToolkit}~\citep{loeffler_2011_et,
  EinsteinToolkit_etal:2022_05}, exploiting the \texttt{Carpet} AMR
driver \citep{Schnetter:2006pg} and the evolution code-suite developed in
Frankfurt consisting of the \texttt{FIL} code for the higher-order
finite-difference solution of the GRMHD equations and of the
\texttt{Antelope} spacetime solver~\citep{Most2019b} for the evolution of
the constraint damping formulation of the Z4 formulation of the Einstein
equations~\citep{Bernuzzi:2009ex, Alic:2011a}. Most of the analysis of
the data presented in the following sections is performed with the
\texttt{kuibit} Python library~\citep{kuibit21}.

In what follows we will present a series of standard and new
implementation tests for a two-moment radiation-transport schemes in
general relativity. The tests aim at asserting the efficacy of the
various aspects of the numerical scheme employed by \FMONEB. It is worth
pointing out that all the tests were performed on a 3D Cartesian grid
even when underlying symmetries are present in the test and that, unless
otherwise stated, all the tests employ a fixed CFL factor of $0.25$.

\subsection{Straight Beam}
\label{ssec:straight_beam}

The first test presented is the so-called ``straight-beam'' test and
consists in setting up a beam of radiation oriented along a single
spatial direction. We therefore set up a grid covering $[0,0.75] \times
[-0.2,0.2] \times [-0.1,0.1]\in \mathbb{R}^3$ with $300\times 160 \times
80$ equidistant points. A radiation beam parallel to the $x-$axis is
injected from the left side of the domain by setting $J=1=H_x$ in the
outer boundary ghostzones of the domain with $x=0$.

Since the spacetime is flat and the background fluid is perfectly
transparent (\ie~we set $\kappa^{\rm er}=Q^{\rm er}=0$), 
the beam is expected to travel straight towards the positive end of the
$x-$axis and at the speed of light. The outcome of this test can be
observed in the left panel of Fig.~\ref{fig:straight_beam}, which reports
the distribution of the normalised radiation energy density at
$t=0.4$. Note that \FMONEA maintains a reasonably focused straight beam,
which propagates at the correct speed as indicated by the expected
position of the beams' front, which is marked with a vertical red
line. Naturally, a small amount of diffusion is present both at the edges
of the beam and at its leading front. This diffusion converges away
with resolution and is the result of the interaction with the surrounding
artificial atmosphere.

\subsection{Curved Beam}

Since \FMONEA is meant to be employed in general-relativistic
simulations, we next consider a test that involves curved spacetimes and
hence non-trivial values for the metric and curvature terms on the
right-hand-side of the radiative-transfer equations. For this reason, we
consider another classical test where a beam of radiation is injected in
a Schwarzschild spacetime to validate the capability of the code of
captured correctly the bending of rays in a curved spacetime~\citep[see
  also][]{Foucart2015a, Radice2022, Weih2020b}. We therefore set up a
grid spanning $[0,5] \times [-0.1,0.1] \times [0,4] \in\mathbb{R}^3$ with
a uniform spacing of $\Delta x=0.05$ in all directions. We then employ a
background fluid that is completely transparent and static around a
unit-mass Schwarzschild black hole place at the origin of the Cartesian
grid. The beam is injected in the region where $x=0$ and $3.0\leq y \leq
3.5$ with $J=1$ and the energy flux set up so that $H^iH_i=J^2$,
$H^{y,z}=-\beta^{y,z}\,J/\alpha$, so that the beam is initially parallel
to the $x-$axis (in this test the CFL factor is set to $0.2$ as required
by a stable Runge-Kutta-3 evolution scheme).

The results of this curved-beam test are reported in the right panel of
Fig.~\ref{fig:straight_beam} and show that \FMONEA is able to reproduce
the bending of radiation geodesics fairly accurately, albeit with the
presence of some non-negligible dissipation. In turn, this leads to a
significant broadening of the flux and consequent reduction of the
intensity of the beam's section when it reaches the $x-$axis at $z=0$.
(the integral of the energy density at $z=0$ is reduced by $50\%$ when
compared with the initial value). Also shown in the figure with dashed
red lines are the two geodesics bounding the initial beam of
radiation. As can be readily observed, the vast majority of the beam's
energy remains enclosed within the analytically correct
trajectories. Overall, this behaviour is in line with what is seen from
other similar codes in the literature when performing this
test~\citep{Foucart2015a, Radice2022, Weih2020b}.

\subsection{Shadow casting}
\label{ssec:shadow}

In the next two tests we consider the interaction of a radiation beam
with an obstacle whose absorption properties are extremely large so that
the radiation will be fully absorbed when interacting with the
obstacle. In turn, this will imply that the radiation field downstream of
the obstacle will be very small and thus the obstacle will effectively
cast a shadow. Both of the tests are performed in a flat spacetime.

\begin{figure}
  \includegraphics[width=\columnwidth]{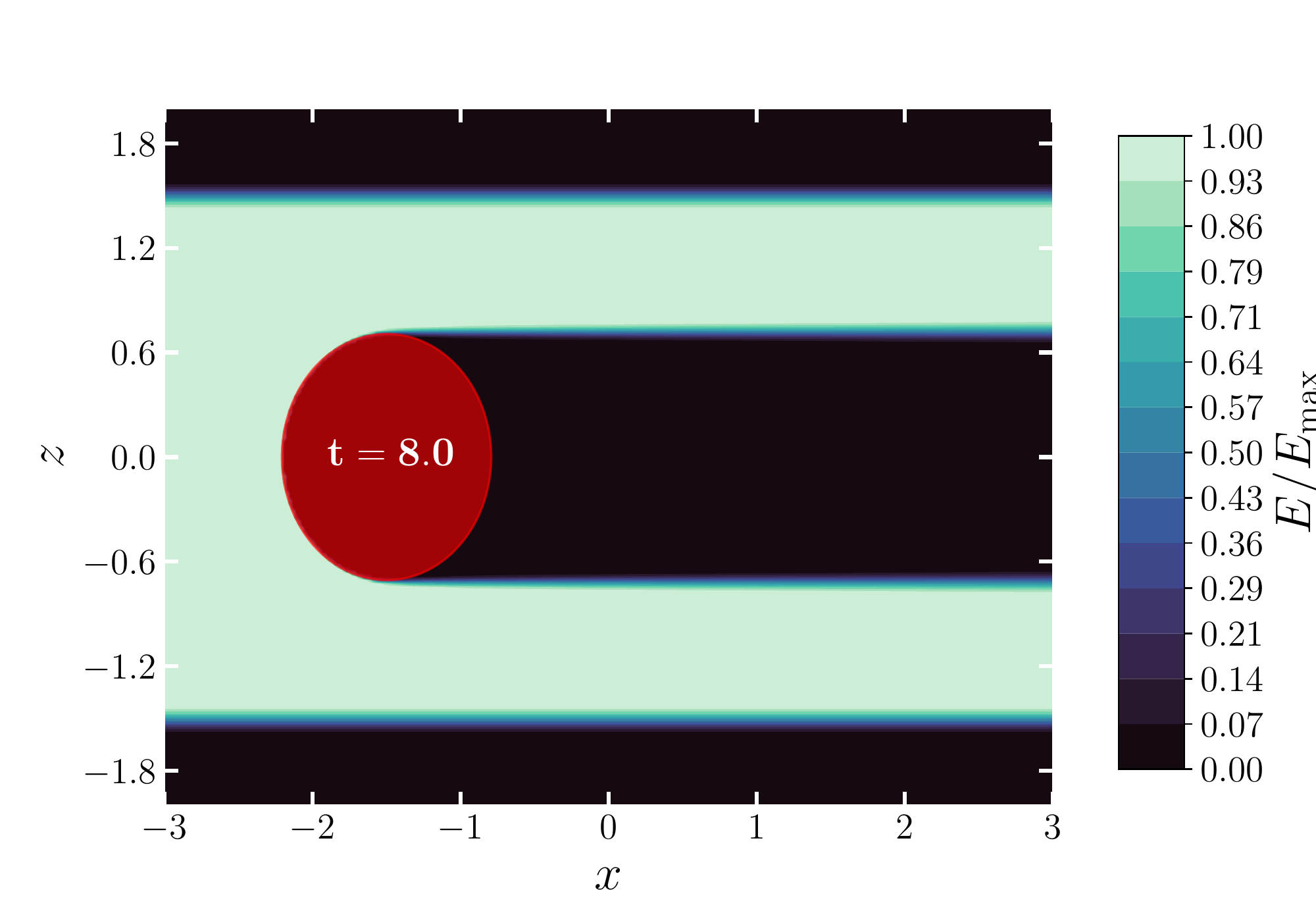}
  \caption{Distribution at $t=8.0$ of the radiation energy density
    normalised to the maximum value for a radiation beam propagating in a
    Minkowski spacetime and impacting an absorbing sphere (shadow-casting
    test). Note that very little diffusion is present downstream of the
    sphere.}
  \label{fig:straight_shadow}
\end{figure}

We first consider a straight beam as the one presented in
Sec.~\ref{ssec:straight_beam} entering the Cartesian domain from the left
boundary. The grid spans $[-4,4] \times [-2.5,2.5] \times
[-2.5,2.5]\in\mathbb{R}^3$ and is covered by $640\times400\times 400$
equidistant points. The spacetime is again a flat one and the domain is
optically thin except for a cylindrical region $\mathcal{A}$ centered at
$x_\mathcal{A}=(-1.5,0,0)$ with a radius of $r_\mathcal{A}=0.5$, where
the opacity to absorption is set to a very large number, \ie $\kappa^{\rm
  er}_a(x\in\mathcal{A})=10^{10}$. The results of the test are presented
in Fig.~\ref{fig:straight_shadow} and clearly show that \FMONEA is able
to produce a sharp shadow with only little diffusion behind or inside of
the high-absorption region.

Next, in order to determine the extent to which the Cartesian grid
affects the beam propagation and absorption onto the obstacle, consider a
scenario in which the radiation field has a spherical
geometry. Therefore, following~\citet{Just2015b} and~\citet{Kuroda2016},
we set up a Cartesian grid spanning $[-7.5,7.5] \times [-7.5,7.5] \times
[-7.5,7.5] \in\mathbb{R}^3$ and covered by $196^3$ equidistant points. We
furthermore set up a refinement level centered at the origin whose radius
we set to $2.5$ and whose refinement factor is taken to be two. This
allows us to check for a potential interference among refinement levels
with the two-moment scheme.

The geometry is again flat and the radiation energy density is set to a
small but positive value everywhere of $E_0 = 10^{-15}$. The radiation
energy flux is initially set to zero on the whole grid. The domain is
optically thin except for a central spherical emitting region
$\mathcal{E}$ of radius $r_\mathcal{E}=1.5$ and an absorbing sphere
$\mathcal{A}$ of radius $r_\mathcal{A}=1$ centered at
$x_\mathcal{A}=(3.5,0)$. We then set the scattering opacity to be zero
across the whole domain, while the absorption opacity is set to follow
the prescription
\begin{equation}
  \label{eq:kappa_a_spherical_shadow}
  \kappa^{\rm er}_a = \begin{cases} 
    10\exp\left[ -\left( 4\sqrt{x^2+y^2}/r_\mathcal{E} \right)^2 \right]
    & x\in\mathcal{E} \\ 
    10  & x\in\mathcal{A}\,.
  \end{cases}
\end{equation} 
The emissivity is set to have an equilibrium energy density $J_{\rm eq} =
10^{-1}$ for $x\in\mathcal{E}$ and to understand the rationale behind
this choice we recall that sources for $\kappa^{\rm er}_s=0$ and $v^i=0$ are [see
  Eqs.~\eqref{eq:sources_eul} and \eqref{eq:M1_G}]
\begin{equation} \label{eq:sources_kappa_s_zero}
  \boldsymbol{\tilde{G}} = \begin{pmatrix}
    \alpha \kappa^{\rm er}_a \left[ {Q^{\rm er}}/{\kappa^{\rm er}_a} - J + H^\mu n_\mu \right] \\
    \\ ~ \\
    -\alpha \kappa^{\rm er}_a \gamma_{\mu i} H^\mu\,,
  \end{pmatrix}
\end{equation}
from which it is apparent that for zero flux $J_{\rm eq} = Q^{\rm er} /
\kappa^{\rm er}_a={\rm const.}$.

\begin{figure*}
  \includegraphics[width=\textwidth]{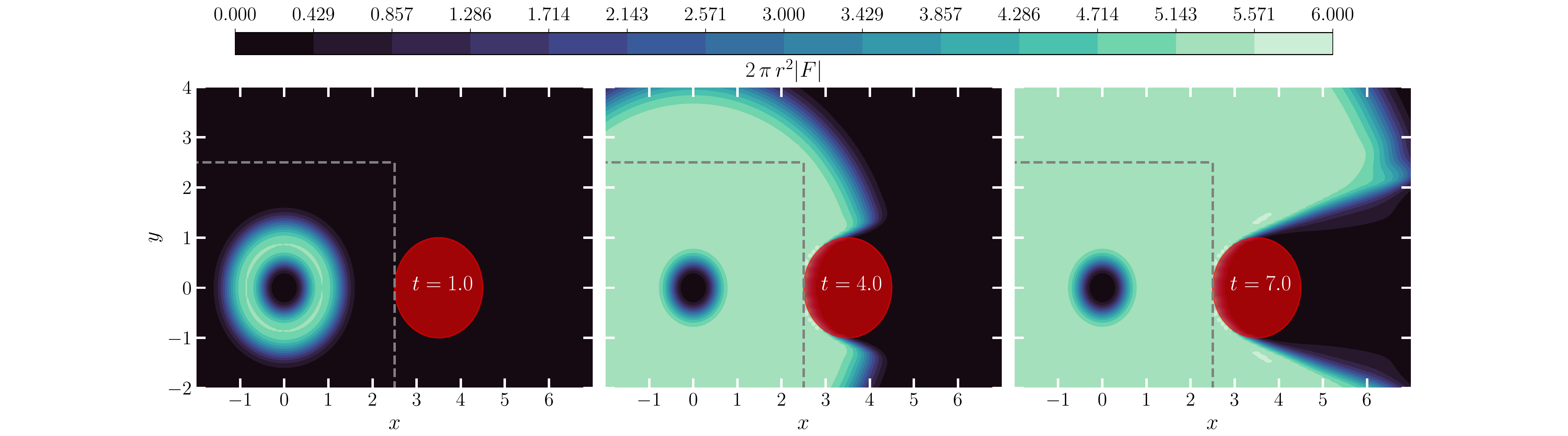}
  \caption{Distributions at different times of the integrated radiation
    flux $2\pi r^2 |F|$ of an expanding ring of radiation propagating in
    a Minkowski spacetime and impacting an absorbing sphere
    (spherical-expansion test). Note that also in this case only a small
    diffusion is present downstream of the sphere despite the propagation
    is not along one of the main coordinates of the numerical grid.}
\label{fig:spherical_shadow}
\end{figure*}

As can be seen from Fig.~\ref{fig:spherical_shadow}, \FMONEA handles this
test-case almost as well as its counterpart where the radiation flux was
aligned with one of the grid principal directions, without excessive
diffusion near the absorbing region and without artefacts spoiling the
spherical symmetry of the simulation due to the Cartesian grid or the
presence of a refinement level (only small oscillations can be seen at $x
\simeq 3.5$). It is worth pointing out that the flux factor is limited to
be smaller than $0.999$ in \FMONEB.

\subsection{Diffusion in a stationary background medium}
\label{ssec:static_diffusion}

\begin{figure*}
  \includegraphics[width=\columnwidth]{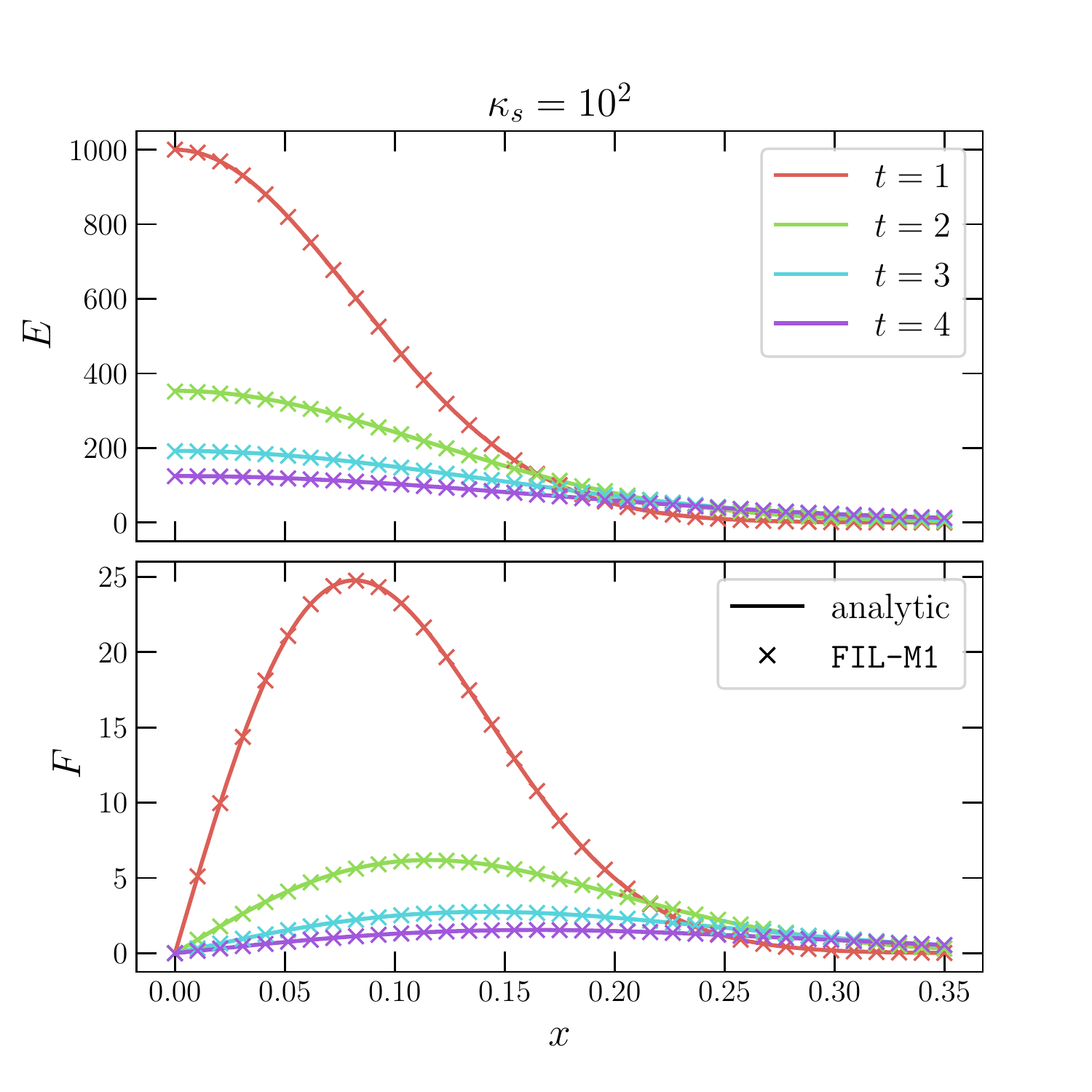}
  \includegraphics[width=\columnwidth]{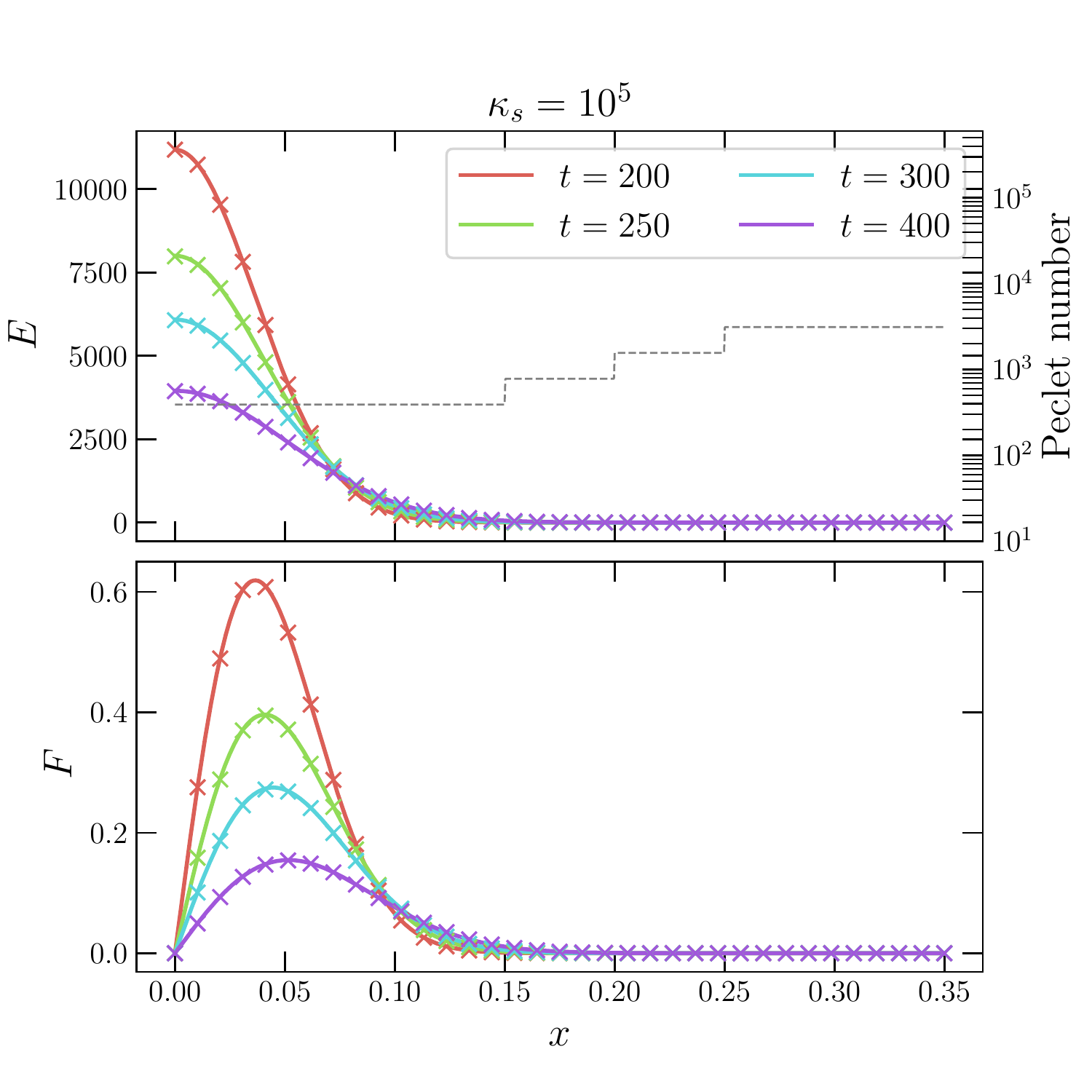}
        \caption{Radial profiles of the energy density (top panels) and
          of the energy flux density (bottom panels) at different times
          of sphere of radiation diffusing in a Minkowski spacetime
          (diffusion-wave test). The left column represents the solution
          for $\kappa^{\rm er}_s=10^2$, whereas the right column for
          $\kappa^{\rm er}_s=10^{5}$. The numerical (analytical)
          solutions are shown with crosses (solid lines) and show a very
          good agreement between the two. Also shown with a dashed gray
          line in the top-right panel is the Plecet number, which is
          clearly always much larger than unity.}
        \label{fig:static_diffusion}
\end{figure*}

The next test probes the effectiveness of \FMONEA in capturing the
asymptotic diffusion limit of the M1 equations. We follow the standard
approach in this type of test~\citep[see, \eg][]{Pons2000, Kuroda2016,
  Weih2020b, Radice2022} and track the evolution of a spherical radiation
wave whose profile is initially a Dirac $\delta-$function at the origin
of the Cartesian grid. Hence, we set up a stationary medium in flat
spacetime with $\kappa^{\rm er}_a=Q^{\rm er}=0$ and $\kappa^{\rm er}_s=(10^2,
10^5)$. Within this scenario, the M1 system reduces to a diffusion
equation for the energy density whose analytic solution is given
by~\citep{Pons2000}
\begin{align}
  &E(r,t) =
  \left(\frac{\kappa^{\rm er}_s}{t}\right)^{d/2}\,\,\exp\left(\frac{-3\kappa^{\rm er}_s
    r^2}{4T}\right)\,, \label{eq:E_diffusion}\\ &F(r,t) =
  \frac{r}{2t}E(r,t)\,, \label{eq:F_diffusion}
\end{align}
where $d$ is the number of dimensions. We therefore layout a 3D Cartesian
domain $[-0.5,0.5] \times [-0.5,0.5] \times [-0.5,0.5]\in\mathbb{R}^3$
covered by $100^3$ points in the case of $\kappa^{\rm er}_s=10^2$ and
$32^3$ points with two nested refinement levels at $x=\pm0.25$ and
$x=\pm0.15$ in the case of $\kappa^{\rm er}_s=10^5$. As a result, the
diffusion-limiter factor $A$ appearing in Eq.~\eqref{eq:F_flux_mod} is
simply given by $A=1$ [see Eq.~\eqref{eq:HLLE_E_mod}], so that the
equations are being solved with the full HLLE solver. In the case of
$\kappa^{\rm er}_s=10^5$, on the other hand, the Peclet number, defined
as the opacity normalised to the grid spacing, \ie ${\rm Pe}:=\kappa^{\rm
  er}_s\,\Delta x$, is ${\rm Pe}\simeq 10^3$ at the center of the grid
and changes with the refinement level. As a result, the equations are
solved with the diffusion part almost completely switched off in the
Riemann solver.

Figure~\ref{fig:static_diffusion} shows the results obtained by \FMONEA
in this test and reports the very good agreement with the analytic
solution, thus validating the ability to properly capture the
asymptotically diffusive limit of the M1 equations.

\subsection{Diffusion in a moving background medium}

\begin{figure}
  \includegraphics[width=0.48\textwidth]{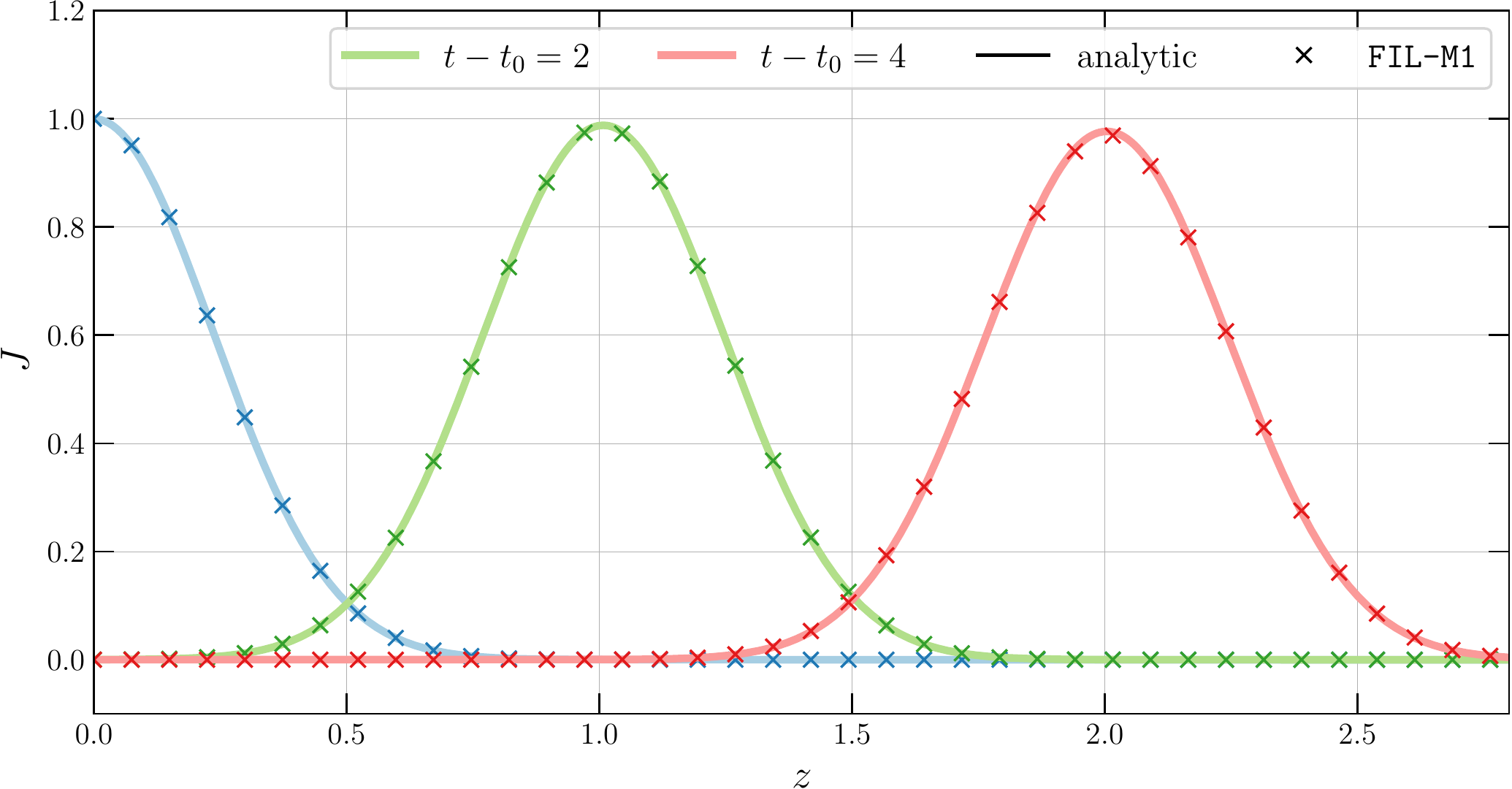}
  \caption{Radial profiles at different retarded times $t-t_0$ of the
    energy density in the frame comoving with the fluid $J$ (moving-fluid
    diffusion test). The numerical (analytical) solutions are shown with
    crosses (solid lines) and show a very good agreement between the
    two. }
  \label{fig:moving_diff}
\end{figure}

Despite the fact that in some of the tests considered so far (in
particular the ones shown in Sec.~\ref{ssec:shadow} and
\ref{ssec:static_diffusion}) the code was confronted with stiff
collisional source terms, none of these has really put the implicit
source-term solver to the test. Indeed, whenever the background is set to
be static ($v^i=0$), the fluid frame moments obviously coincide with the
Eulerian ones, and the source terms become linear functions of the
evolved variables. In particular, the initial guess used by the solver
Eq.~\eqref{eq:solver_initguess} are already the solution. We therefore
set out to perform the test first introduced by~\citet{Radice2022}, where
a diffusing radiation wave is evolved on a moving background medium. In
particular, we set the scattering opacity $\kappa^{\rm er}_s=10^3$, a
background velocity $v^x=0.5$ and initialise the radiation energy density
in the Eulerian frame to simply be
\begin{equation}
  \label{eq:ID_moving_diff}
	E = \exp(-9x^2)\,.
\end{equation}
We then use Eq.~\eqref{eq:F_Eulerian_thick} with $H_\alpha=0$ to
initialise the flux assuming a completely optically-thick medium. As
fiducial solution to compare to, we employ Eq.~\eqref{eq:E_diffusion}
translated at a velocity $v^x=0.5$, with $d=1$ due to the slab geometry
of the problem and with a diffusion coefficient appropriately scaled to
get Eq.~\eqref{eq:ID_moving_diff} as initial condition. For this test we
employ a grid covering $I = [-3,3]$ in the $z-$axis and spanning only a
few points in the other two directions. The grid is covered by $600$
points on the $z$-axis, leading to a spacing of $\Delta x=0.01$ code
units. The numerical solution obtained by \FMONEA is shown in
Fig.~\ref{fig:moving_diff}, and reproduces the fiducial solution within
the expected error level. The importance of this test is that it probes
the capability of the code of dealing with stiff sources in a moving
medium where the nonlinear solver is strictly required in order to
converge to the true solution. Moreover, it provides an additional
validation of the flux formulation in the diffusive limit.

\subsection{Uniformly radiating-sphere test}

The final test we consider in a flat spacetime is represented by the
so-called uniformly radiating-sphere test and consists in a spherical
region $\mathcal{E}$ of radius $R_\mathcal{E}$ centered around the origin
of the coordinate system and that is allowed to emit and absorb
radiation. The sphere is surrounded by vacuum and the opacity and
emissivity are therefore defined as 
\begin{equation*}
	\begin{cases}
		Q^{\rm er} := \kappa^{\rm er}_a = 1 \qquad x \in \mathcal{E}\,, \\
		Q^{\rm er} := \kappa^{\rm er}_a = 0 \qquad \text{elsewhere}\,.
	\end{cases}
\end{equation*}
While this test can be thought of as a very crude model for an
astrophysical spherical source of radiation, its important lays in
serving as a test of the collisional source terms when the absorption
coefficient is nonzero. This problem admits an analytic solution where
the distribution function is given by~\citep{Pons2000, Murchikova2017}
\begin{equation}
  \label{eq:BoltzDistr_RadSphere}
  f(r,\mu) = \eta \Bigg[ 1 - \exp\left({-\kappa^{\rm er}_a \,
      R_\mathcal{E}\, s(r,\mu)}\right)\Bigg]\,,
\end{equation}
with 
\begin{equation*}
  s(r,\mu) = 
  \begin{cases} 
    {r\,\mu}/{R_\mathcal{E}} + g(r,\mu) & r< R_\mathcal{E} \,\, {\rm and}
    \,\, -1\leq\mu\leq 1\,,\\ 2 g(r,\mu) & r\geq R_\mathcal{E} \,\, {\rm
      and} \,\, \sqrt{1-\left({R_\mathcal{E}}/{r}\right)^2} \leq \mu \leq
    1\,, \\ 0 & \text{elsewhere}\,,
  \end{cases}
\end{equation*}
and 
\begin{equation*}
  g(r,\mu) := \sqrt{1-\left(\frac{r}{R_\mathcal{E}}\right)^2\,(1-\mu^2)}\,.
\end{equation*}
For this test we extend the domain to $4\,R_{\mathcal{E}}$ in all
directions with a resolution of $\Delta x=0.0125$ and employ reflection
symmetries across all coordinate planes to reduce computational costs.

\begin{figure}
  \includegraphics[width=\columnwidth]{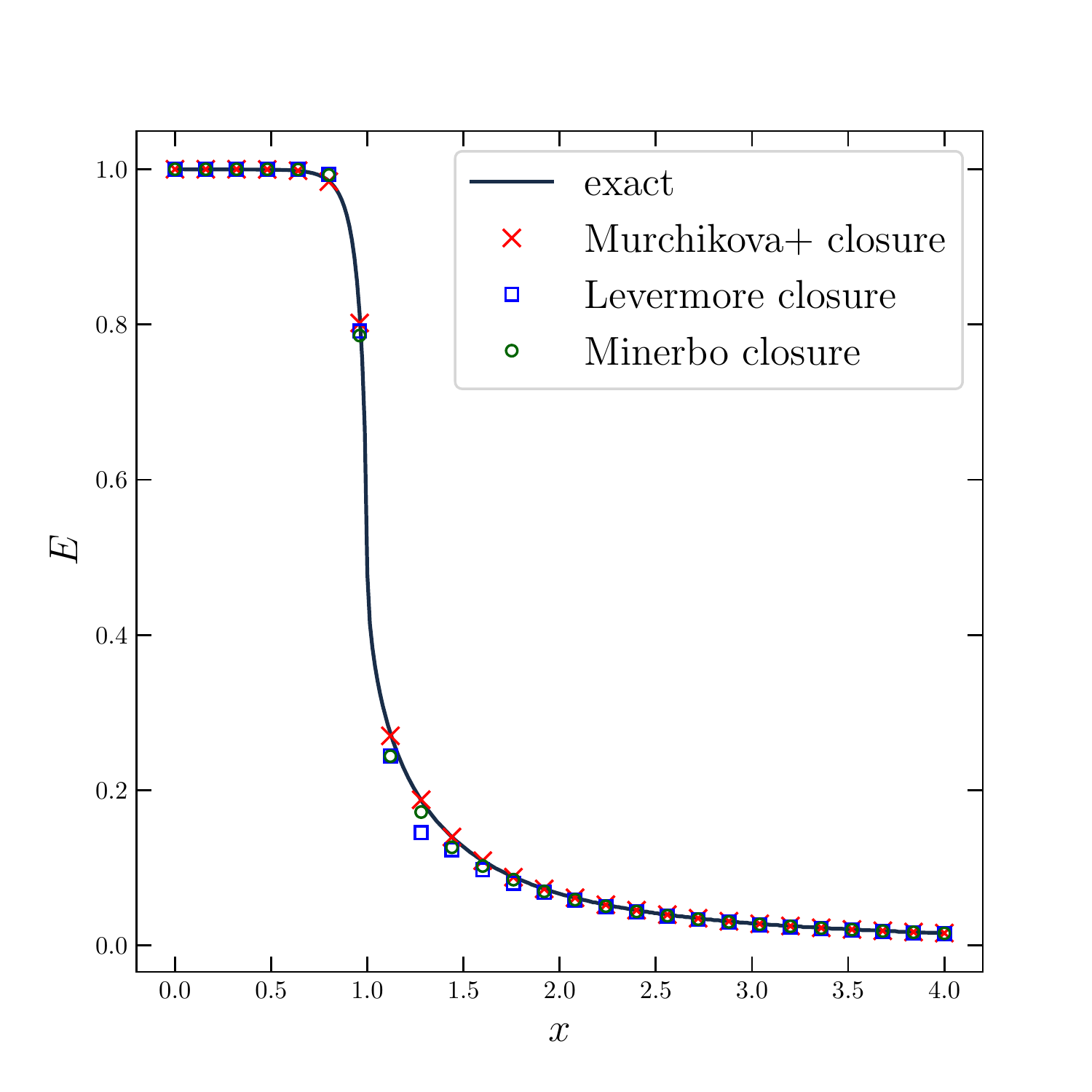}
  \caption{Radial profiles of the energy density $E$ for
    a uniform sphere of radiation (uniform-sphere
    test). The results for different closures (Murchikova+, Levermore and
    Minerbo) are shown with symbols, while the exact solution is shown
    with a black solid line. }
  \label{fig:radsphere}
\end{figure}

The results obtained by \FMONEA are shown in Fig.~\ref{fig:radsphere}
using three different closures: the Minerbo closure~\citep{Minerbo1978},
which is the default for \FMONEB, the Levermore
closure~\citep{Levermore1984}, which consists in always assuming an
optically-thick regime in the fluid frame, and with the fit to the
analytic closure proposed by~\citet{Murchikova2017}. As can be seen, the
Minerbo closure reproduces the result quite well although it does not
converge to the analytic solution. On the other hand, the fit to the
analytic closure of~\citet{Murchikova2017} is accurate to within $1\%$
for this specific problem and \FMONEA can be seen to accurately reproduce
the analytic result with this choice. These results emphasise how the
choice of a closure that is most appropriate for the problem at hand can
considerably improve the accuracy of the M1 approximation. It is
therefore desirable when developing an M1 code to allow by construction
for the use of arbitrary closures, as done in \FMONEA.

\section{Neutron-star simulations}
\label{sec:convtest}

\begin{figure*}
  \includegraphics[width=.9\textwidth]{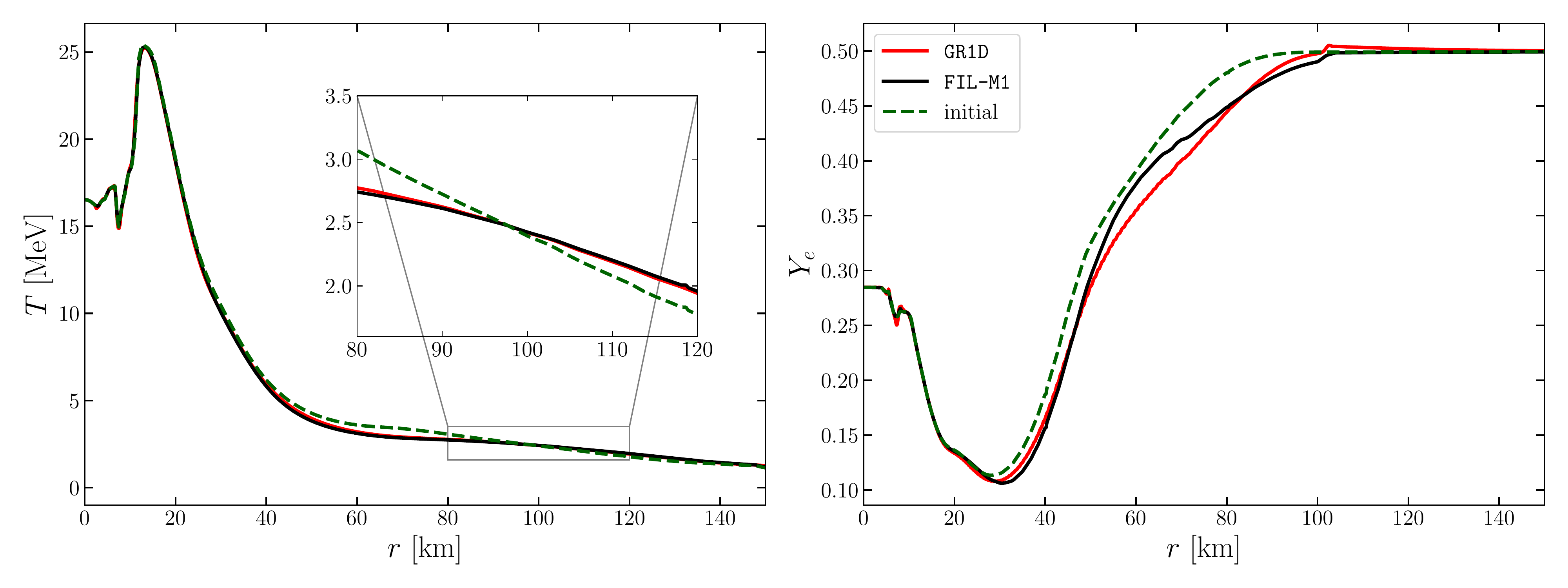}
  \caption{Radial profiles of the temperature (left panel) and of the
    electron fraction (right panel) in the evolution of a core-collapse
    supernova remnant as computed by \FMONEA (black solid line) and by
    the \texttt{GR1D} code (red solid line). The curves refer to the last
    available snapshot at $t - t_\text{ini.} \simeq 8\, \mathrm{ms}$,
    while the dashed line refers to the initial solution at
    $t_\text{ini.}-t_\text{bounce}\simeq 150\, \mathrm{ms}$. Note that
    agreement in the temperature is within the expected error levels and,
    more importantly, the ``gain region'' , where neutrino absorption
    exceed emission and net heating occurs, starts at $r\gtrsim
    100\mathrm{km}$ and is properly described by the grey scheme of
    \FMONEB.}
  \label{fig:CCSN_T_ye}
\end{figure*}

The next series of tests is rather different as it meant to assess the
ability and self-convergence properties of \FMONEA in providing accurate
solutions under the physical conditions that are expected to be present
in realistic simulations involving neutron stars and where neutrino
transport plays an important role. For this purpose, we consider three
different tests. First, and following previous
literature~\citep{Foucart2015a, Just2015b, Foucart2020montecarlo}, we
compare the results obtained by \FMONEA with those provided by the
open-source energy-dependent M1 code \texttt{GR1D} corresponding to a
proto neutron star resulting from a core-collapse supernova
(Sec.~\ref{sec:CCSN}). This test will show that the results obtained by
our code in a realistic context and when including all source terms and
coupling to the astrophysical plasma are consistent with well-tested
implementations of similar schemes. Secondly, we probe the
self-convergence of \FMONEA in the context of a simple oscillating
nonrotating star (Sec.~\ref{sec:TOV}). Finally, we discuss in detail the
simulation of the head-on collision of two neutron stars
(Sec.~\ref{sec:HEADON}), whose results will be made publicly available so
as to providing a new multidimensional and non-trivial test for future M1
implementations.

\subsection{Core-collapse supernova test}
\label{sec:CCSN}

For this test we simulate the collapse of a $20 M_{\odot}$
progenitor~\citep{Woosley06} employing the tabulated DD2
EOS~\citep{Hempel2009} with the publicly available \texttt{GR1D}
code~\citep{Oconnor2015}. In particular, we import the (1D) radial
solution obtained by GR1D after core bounce into the 3D Cartesian grid
employed my \FMONEA with reflection symmetries across all coordinate
planes. Furthermore, we add five static mesh refinement boxes at radii
$r_{^{\rm RL}}=\{ 40, 80, 120, 160 \}\, M_{\odot}$, so that the outermost
box has an outer boundary at $x_{\rm max}=208\,M_{\odot}$ is covered by
$26\times 26 \times 26$ points leading to a finest spacing of $\Delta
x_{5} = 0.5\,M_{\odot} \simeq 740\,\mathrm{m}$. On the other hand, the
(radial) grid in \texttt{GR1D} covers the domain $[0,13]\,M_{\odot}$ with
a uniform spacing $\Delta x_{u}=0.2\,M_{\odot}$ and extends outwards with
a logarithmically spaced grid. The import is performed with a simple
linear interpolation. The weak rates in \texttt{GR1D} are chosen to be as
similar as possible as those in \FMONEB. In particular, we enable the
re-computation of the emission coefficients from the absorption opacities
using of Kirchhoff's law and we deactivate in \FMONEA the calculation
plasmon-decay emissivity since this is not supported by the the
neutrino-interaction library employed by \texttt{GR1D}, namely,
\texttt{NuLib}~\citep[][]{Sullivan_2015}.

The systems of radiative-transfer equations are then evolved in both
\texttt{GR1D} and \FMONEA from $t_\text{ini.}-t_\text{bounce}\simeq
0.15\, \mathrm{s}$, and up to $t_\text{fin.}\simeq 8\, \mathrm{ms}$,
keeping the spacetime fixed in both codes and updating only the
hydrodynamical variables in \FMONEA through the backreaction of
neutrinos. This is done for two different reasons. First, because the 3D
Cartesian grid cannot cover the whole domain that would allow a
self-consistent hydrodynamical evolution without significant
computational overhead. Second, because we are only interested in
comparing the radiation-transfer implementations across the two codes.

\begin{figure*}
  \includegraphics[width=0.32\textwidth]{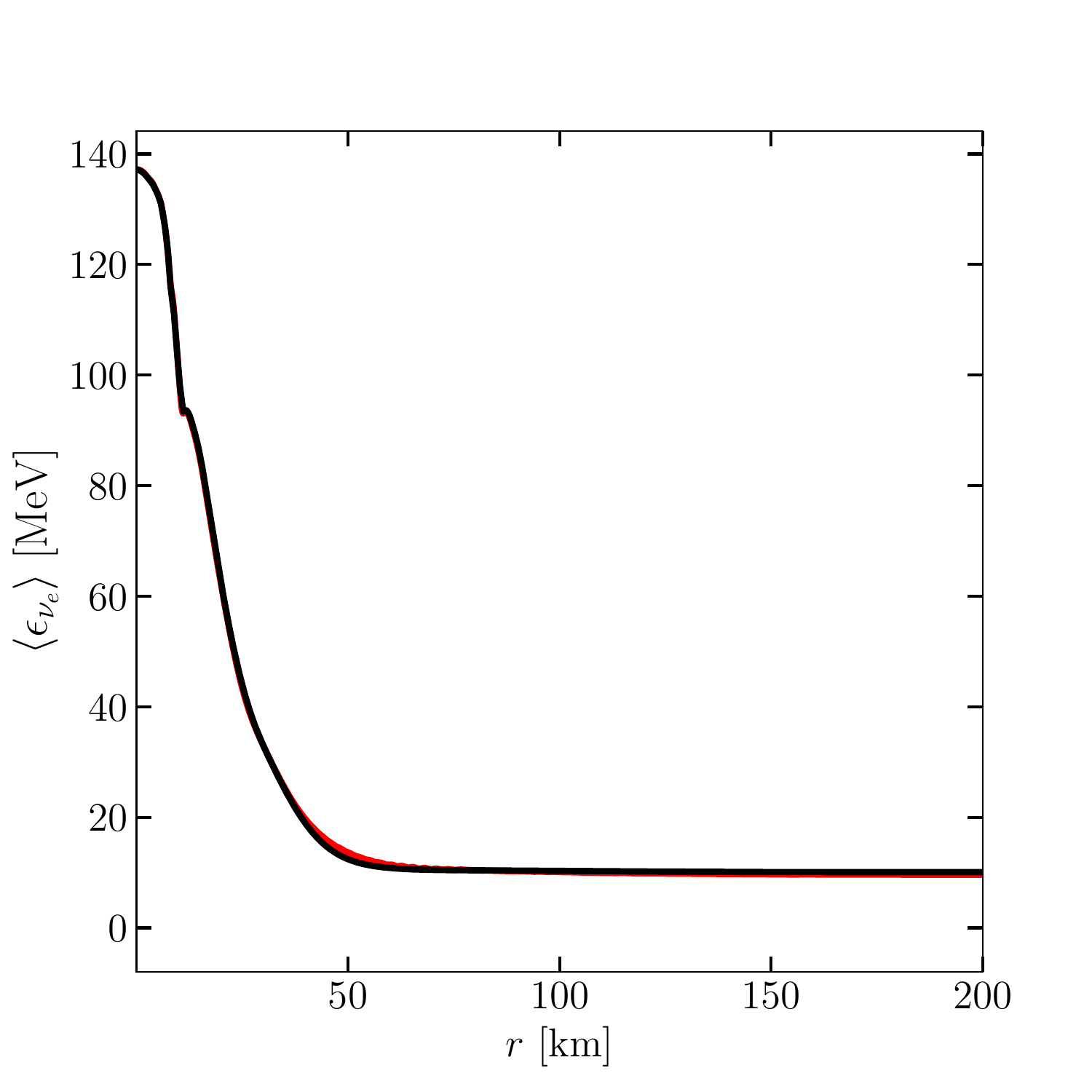}
  \hskip 0.25cm
  \includegraphics[width=0.32\textwidth]{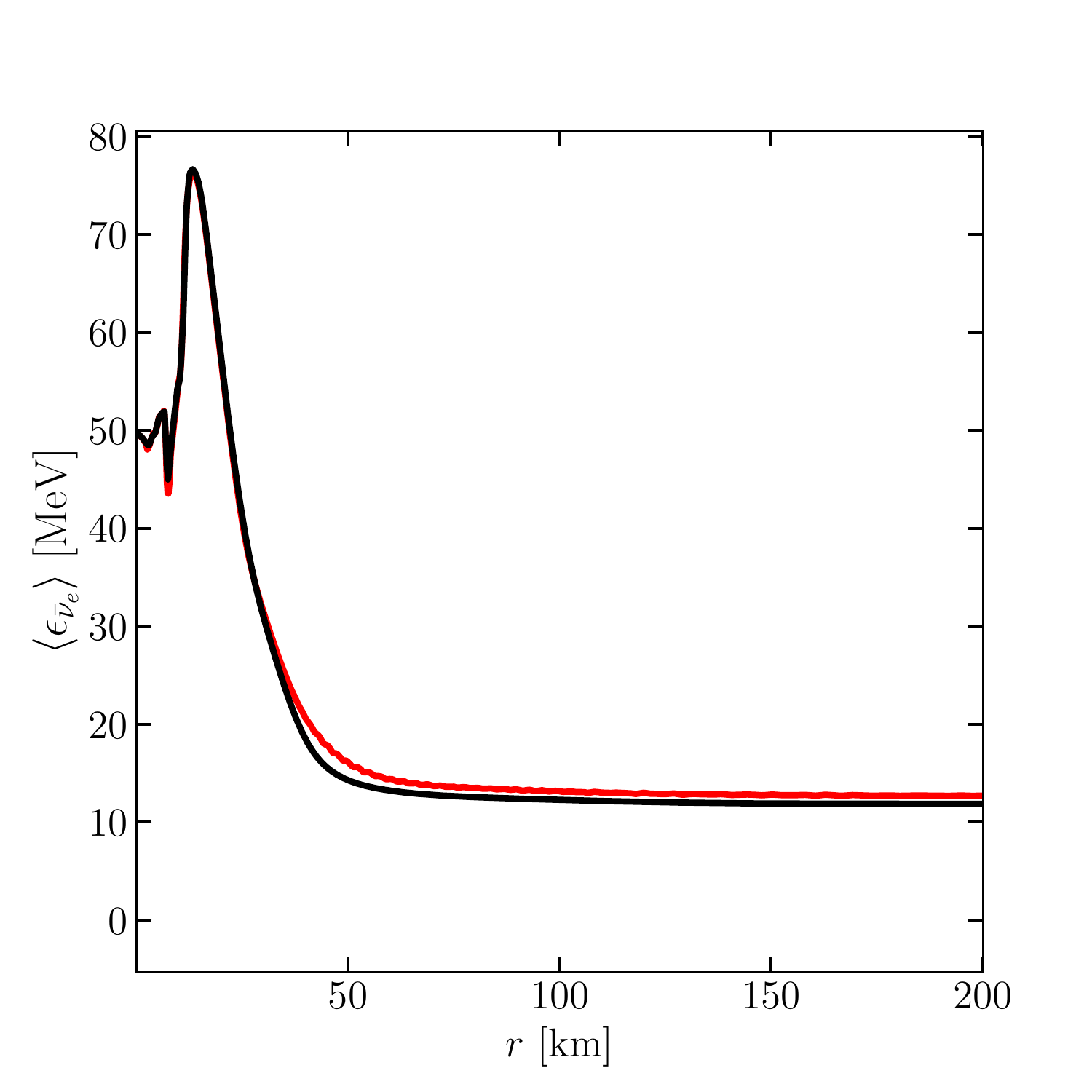}
  \hskip 0.25cm
  \includegraphics[width=0.32\textwidth]{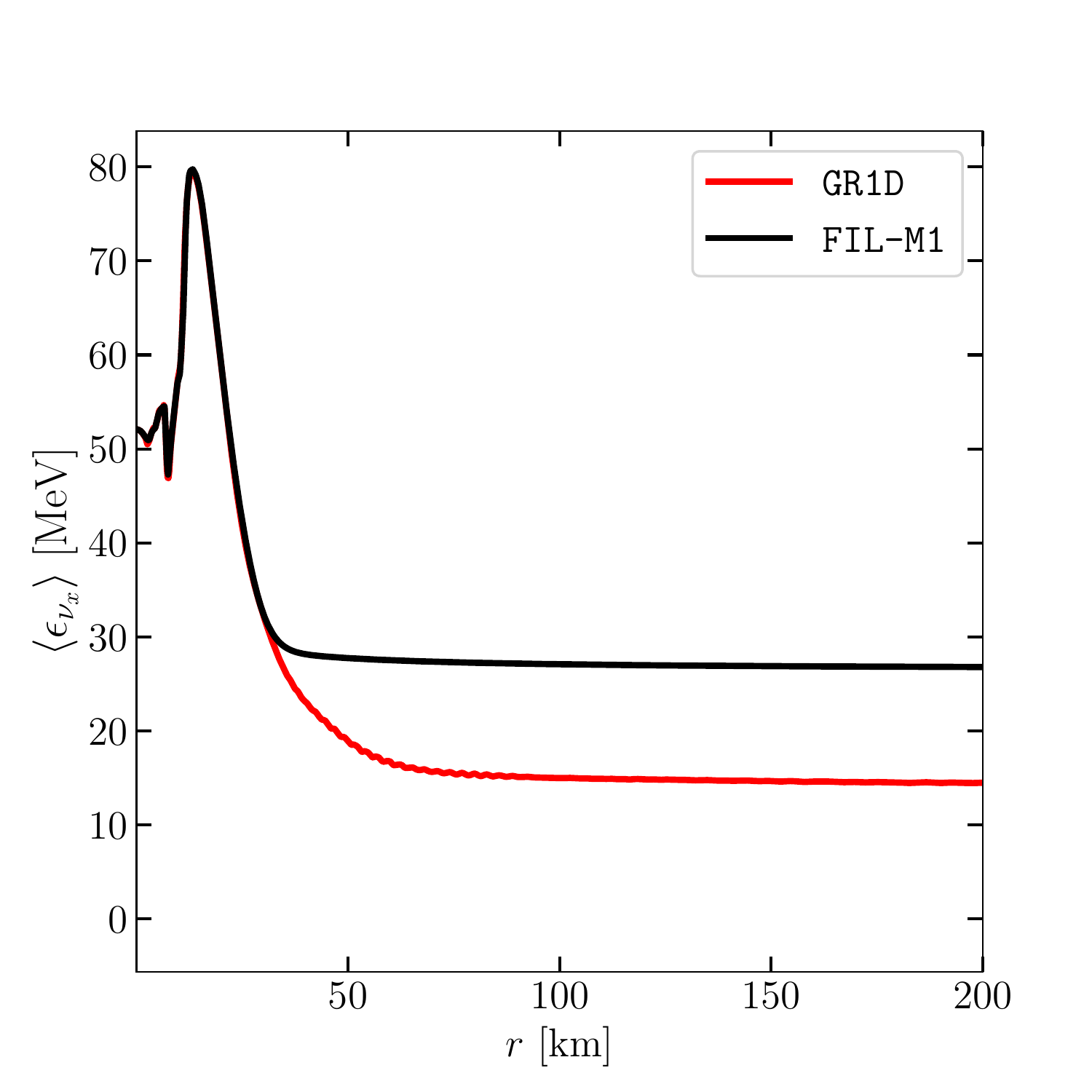}
  \includegraphics[width=0.32\textwidth]{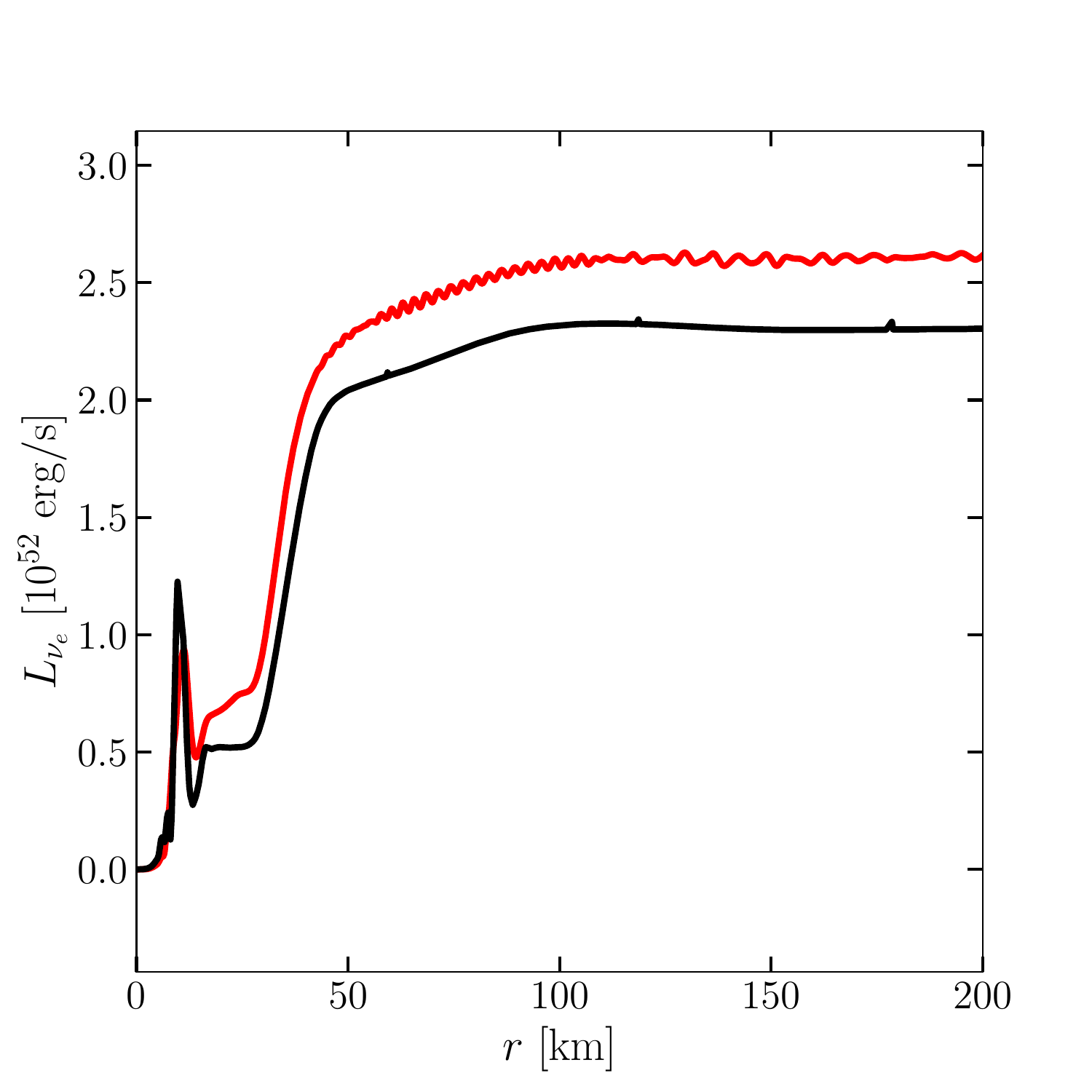}
  \hskip 0.25cm
  \includegraphics[width=0.32\textwidth]{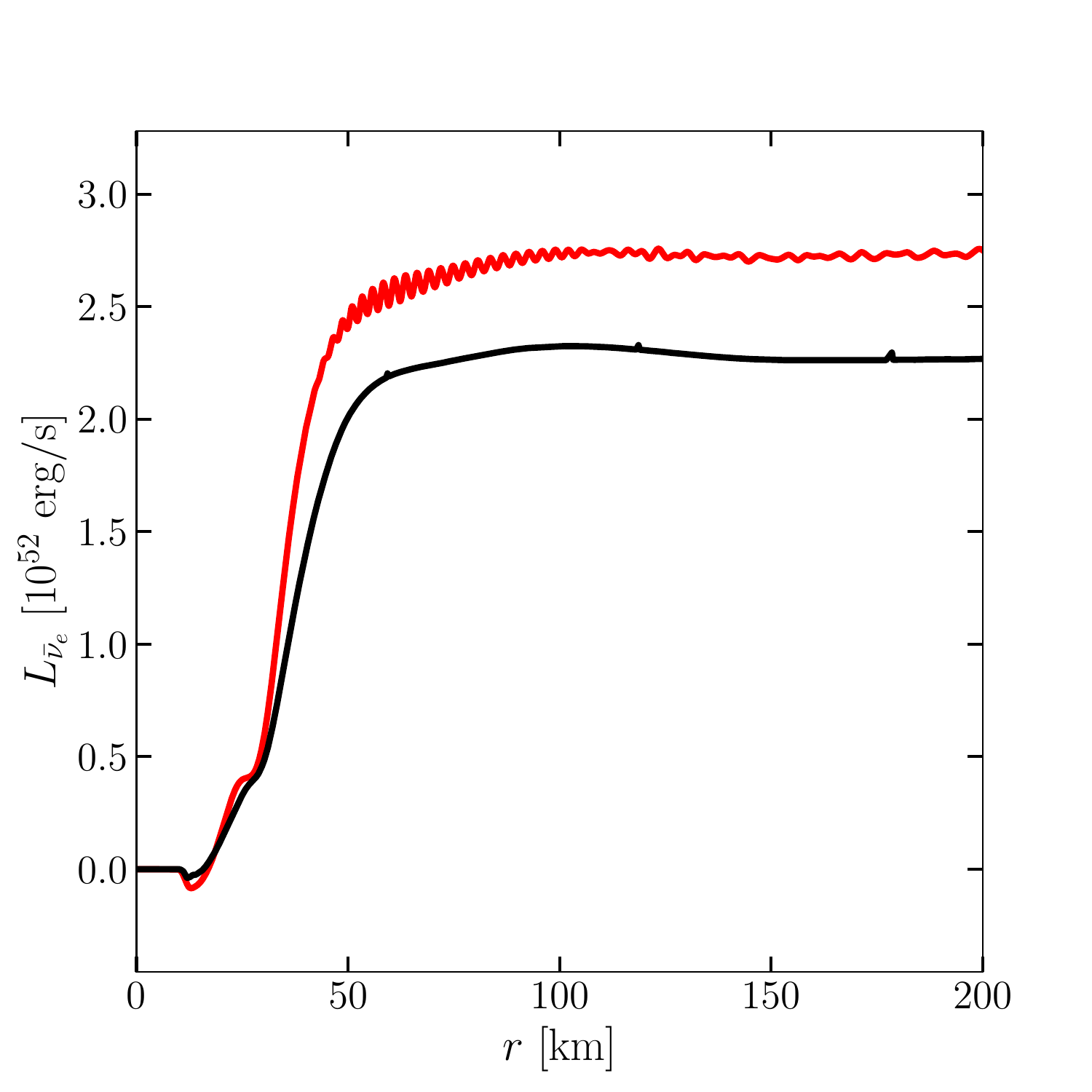}
  \hskip 0.25cm
  \includegraphics[width=0.32\textwidth]{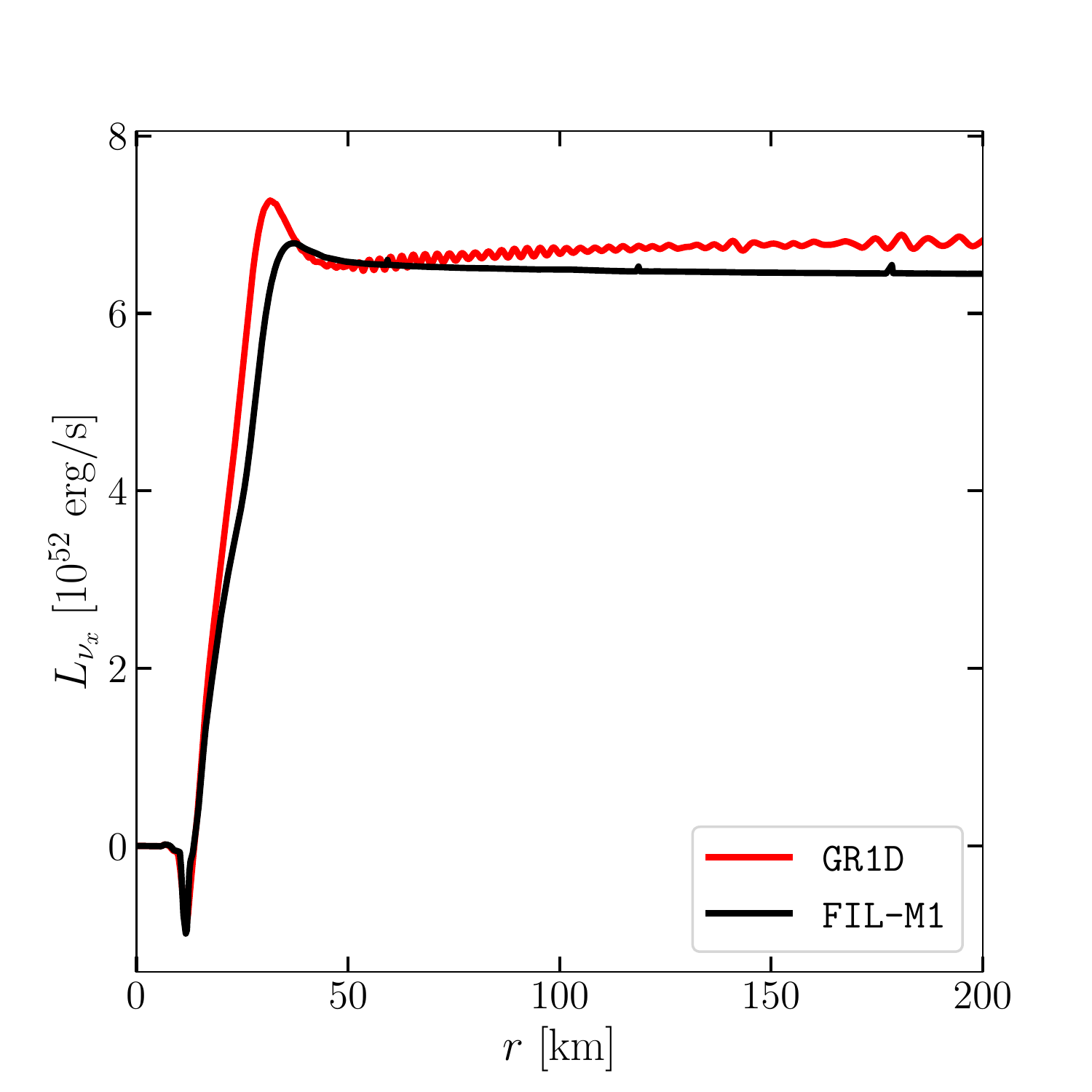}
  \caption{\textit{Top row:} Average neutrino-energy profiles for the
    three neutrino species in the evolution of a core-collapse supernova
    remnant as computed by \FMONEA (red solid line) and by the
    \texttt{GR1D} code (red solid line); the curves refer to the last
    available snapshot at $t_\text{fin.}\simeq 8\, \mathrm{ms}$. Note
    that electron-flavour neutrinos show an excellent agreement across
    the whole domain, whereas heavy-lepton neutrinos suffer from a
    systematic error in the optically-thin region. \textit{Bottom row:}
    The same as above but for the neutrino luminosities
    profiles.}
  \label{fig:CCSN_eps}
\end{figure*}

The comparison of the results obtained by the two codes is shown in
Fig.~\ref{fig:CCSN_T_ye}, with the left panel referring to the
temperature and the right one to the electron fraction. Clearly, the
agreement in the temperatures is remarkable, while the electron fraction
suffers from larger uncertainties. This is to be expected since the
solution from \FMONEA is energy integrated (grey) while that in
\texttt{GR1D} takes into account twelve energy bins.  Overall, the
results reported in Fig.~\ref{fig:CCSN_T_ye} are comparable with those
shown in similar code comparisons~\citep{Foucart2015a}. The overall
consistency between the two radiative-transfer solutions is confirmed
also by the panels in the top row of Fig.~\ref{fig:CCSN_eps}, which show
the profile of the neutrinos average energies for the three neutrino
species. The profiles show very good agreements in the case of
electron-flavour neutrinos, while a systematic overestimation
characterises the effective average energy of heavy neutrinos and was
already reported by~\citet{Foucart2016a} (the case that can be directly
compared to \FMONEA is the one reported as $\beta=\infty$ in Tab.~(1) of
the reference). It is worth noting that the
agreement shown in the average energies is indeed remarkable since all
information about the neutrino energy spectrum is lost in a grey M1
scheme. We believe that the reason why \FMONEA is able to reproduce the
results from the energy-dependent transport code is that it evolves an
additional equation for the number density of neutrinos, namely
Eq.~\eqref{eq:M1_N}. In particular, the correction to absorption and
scattering rates detailed in Eq.~\eqref{eq:epscorr} was already shown in \citet{Foucart2015a}
to be crucial in allowing a grey M1 scheme to capture the features of the
temperature profile in this test, and the average energy of neutrinos
contained in this correction can be estimated more accurately through the
additional information provided by the evolved neutrino number density.
Finally, we report in the bottom row of Fig.~\ref{fig:CCSN_eps} the neutrino
luminosities defined as 
\begin{equation}
L_{\nu_*}(r) := \int_{\Sigma}
d\Sigma^i F^{(\nu)}_{i}\,,
\end{equation}
where $\nu_*$ stands for either $\nu_e$, $\bar{\nu}_e$, or $\nu_x$,
$\Sigma$ is a coordinate 2-sphere placed of radius $r$
and $d\Sigma^i$ its oriented unit normal. Also in
this case, the results presented in lower panels of
Fig.~\ref{fig:CCSN_eps} are comparable with those reported in the
literature and validate the ability of \FMONEA to describe a non-trivial
and dynamical radiative-transfer problem.

\subsection{Self-convergence of a hot nonrotating star}
\label{sec:TOV}

We next consider a very important test that has never been performed in previous
implementations of an M1 scheme, namely establishing the self-convergence
properties of \FMONEA in a fully coupled system in which both the
Einstein and the GRMHD equations are evolved. While this test should be
straightforward in principle, it is extremely challenging in
practice. First, applying such a test to a full binary system would be
computationally prohibitive as the costs associated with a series of
simulations, two of which need to be at resolutions sufficiently
high to show a convergent postmerger behaviour, would easily exhaust the
computational resources typically available in research groups. Second,
these simulations would require sufficiently long times after the merger
to allow the codes to recover from the low convergence order produced by
the large shocks at merger~\citep[see][for a discussion]{Radice2013b,
  Most2019b}. Because of this, it is not surprising that rigorous
self-convergence tests for simulations of binary mergers have not been
performed before in previous implementations of an M1
scheme~\citep[however, see][for self-convergence tests on simpler
  setups]{Radice2022}. 

In view of these considerations we validate the convergence properties of
a simpler scenario, namely, the neutrino luminosity from a spherically
symmetric hot neutron star described by a temperature and
composition-dependent EOS. It should be stressed that while simpler, this
test remains challenging for at least three different reasons. First,
being hot, the star does not have a background equilibrium which it can
converge to as the resolution is increased~\citep{Font02c}; as a result,
although in a convergent regime, the star will exhibit a secular
behaviour as it cools and expands. Second, the inevitable oscillations
that are triggered in the star at the initial time will lead to periodic
weak processes and neutrino emissions that will introduce short-timescale
oscillations in the overall behaviour. Finally, the quantity whose
convergence properties we want to show, namely, the neutrino luminosity,
is a derived quantity and not a conserved one. As such, it will be
affected by a number of other numerical operations (most importantly
root-findings and interpolations) that inevitably will impact the
convergence order. Notwithstanding these considerations, we will show
below that \FMONEA is able to provide a converged solution for this
scenario.

For our test we consider a nonrotating stellar model constructed from the
isentropic slice with entropy per baryon $s=1\,k_{\rm B}$ of the DD2
EOS~\citep{Hempel2009} and with a central density of
$\rho_c=7.41\times10^{14}\,\mathrm{g/cm}^3$; the star has an ADM mass of
$M_{^{\rm TOV}} = 1.98\,M_\odot$ and a central temperature of the order
of $\sim 30\,\mathrm{MeV}$. The star is then evolved with three different
resolutions differing from each other by factors of two and
respectively of $\Delta x_{^{\rm LR}} \simeq 531.6\, {\rm m}$ (low
resolution), $\Delta x_{^{\rm MR}} \simeq 265.8\,\mathrm{m}$ (medium
resolution), and with $\Delta x_{^{\rm HR}} \simeq 132.9 \, {\rm m}$
(high resolution). In all cases, we consider a fixed mesh four refinement
levels, where the coarsest grid extends to $\sim 380\, \mathrm{km}$ in
each direction. The neutrino luminosities are extracted at a coordinate
sphere of radius $R \simeq 150\,\mathrm{km}$, which is large enough to
guarantee that no spurious diffused material will ever reach the
detector.

\begin{figure*}
  \includegraphics[width=\textwidth]{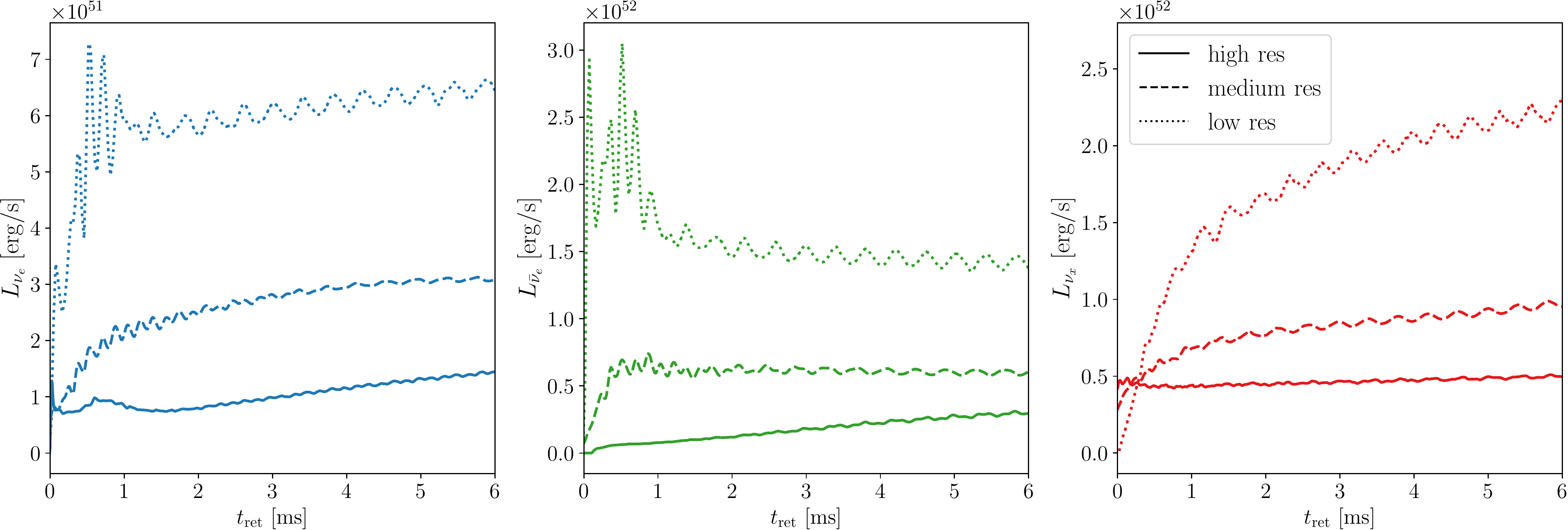}
  \caption{Neutrino luminosities of the three neutrino species considered
    and produce by a hot and radiating nonrotating star. The luminosities
    are shown as a function of the retarded time at the detector and for
    the three different resolutions (high, medium and low) employed in
    the self-convergence test. Note that the strong dependence of the
    location of the neutrinosphere induces changes of up to a factor five
    between the low and high-resolution simulations.}
  \label{fig:TOV_lum}
\end{figure*}

\begin{figure}
  \includegraphics[width=\columnwidth]{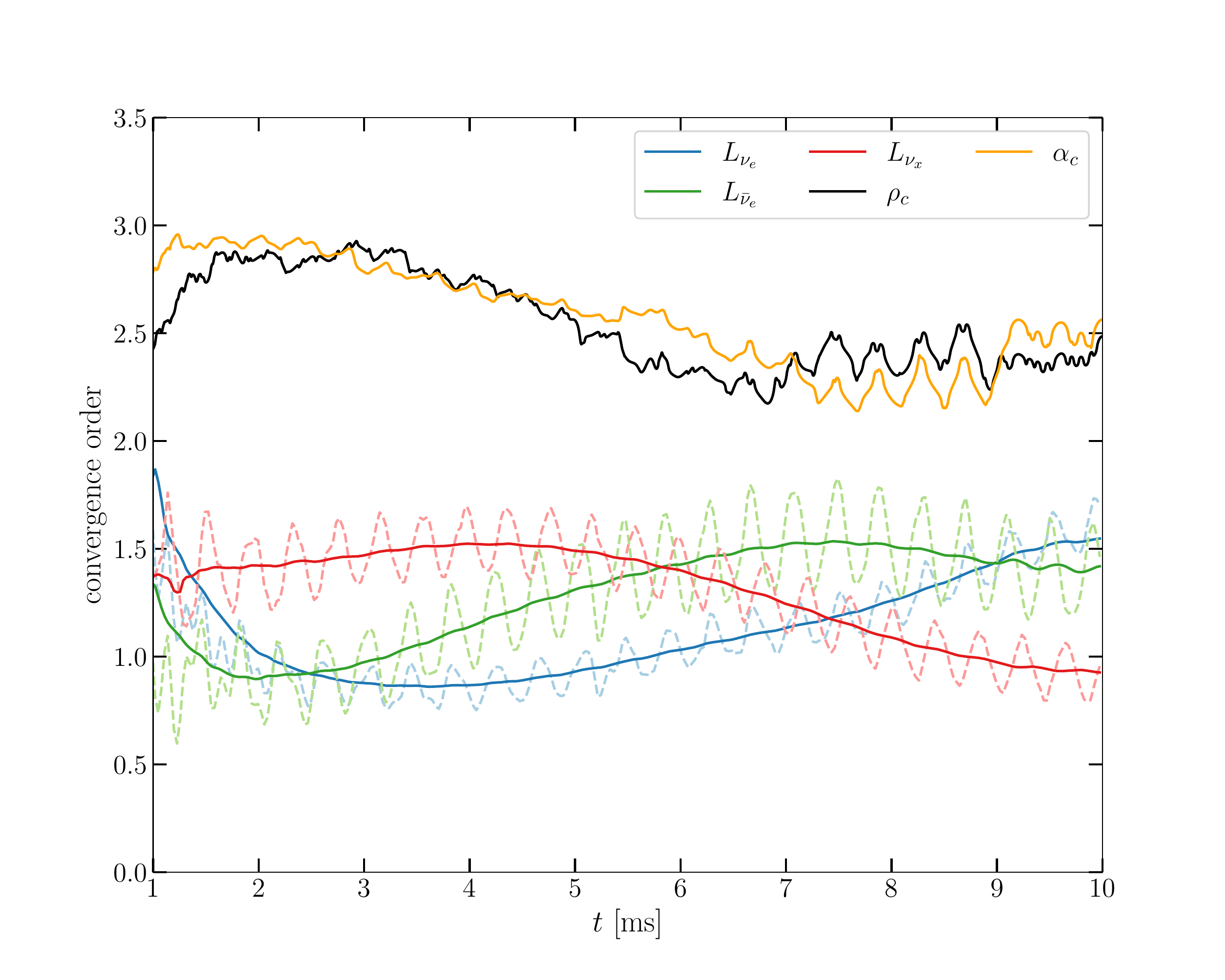}
  \caption{Convergence order for the neutrino luminosities shown in
    Fig.~\ref{fig:TOV_lum}. The thick lines are calculated as a moving
    average with a window of width of $0.2\,\mathrm{ms}$, whereas the
    unfiltered data is shown with the dashed lines of the same
    colour. Also reported are the convergence orders on the pointwise
    values of the central rest-mass density (black solid line) and lapse
    function (orange solid line). The lower convergence order of the
    luminosities is due to the intrinsic lower order of the numerical
    scheme and the strong dependence of the neutrinosphere on the stellar
    temperature near the stellar surface.}
  \label{fig:TOV_conv}
\end{figure}

The luminosities of each neutrino species for each resolution is shown in
Fig.~\ref{fig:TOV_lum}, where it is already clear that the code displays
a very consistent behaviour across resolutions for all of the three
neutrino species, although the strong dependence of the location of the
neutrinosphere induces changes of up to a factor five between the low and
high-resolution simulations (see discussion below). We then compute the
convergence order for a given observable $\boldsymbol{O}$ computed at
resolutions $h,\ell,k$ as~\citep{Rezzolla_book:2013}
\begin{equation}
  \label{eq:conv_order}
  p = \frac{1}{\log(\gamma)}\,\log \left( \frac{ \left|\boldsymbol{O}^{(h)} -
    \boldsymbol{O}^{(\ell)}\right| }{ \left|\boldsymbol{O}^{(\ell)} -
    \boldsymbol{O}^{(k)}\right| } \right)\,,  
\end{equation}
where $h<\ell<k$ and $h=\ell/\gamma$, $\ell=k/\gamma$. In using
expression~\eqref{eq:conv_order} it is of course assumed that the
truncation error on the observable $\boldsymbol{O}$ can be expressed as a
simple power-law with index $p$, which is only true for grid spacings
that are small enough to make the sub-leading terms in the error
expansion negligible (\ie~in the so-called ``convergence regime''). The
convergence order is reported in Fig.~\ref{fig:TOV_conv} for each
neutrino species, showing with dashed lines the instantaneous convergence
order and with solid lines of the same colour the time-average
corresponding order employing a uniform moving filter of
$0.2\,\mathrm{ms}$. It is worth remarking that we compute all convergence
orders using Eq.~\eqref{eq:conv_order} and that in the case of neutrino
luminosities $\boldsymbol{O}$ is intended as the timeseries of the
observable, whereas for the lapse and rest-mass density $\boldsymbol{O}$
indicates the pointwise value of the field evaluated at the center of the
grid.

\begin{figure}
  \includegraphics[width=\columnwidth]{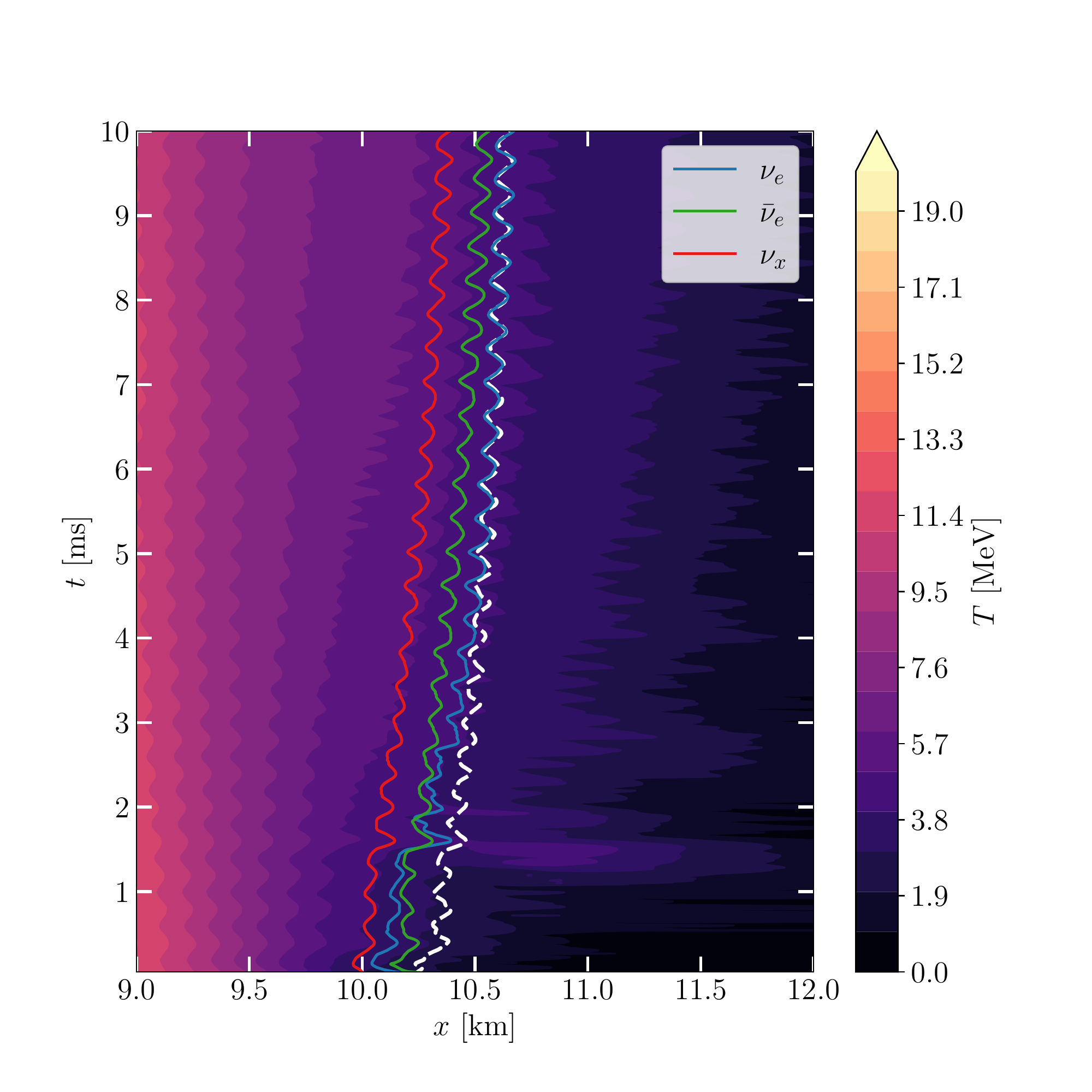}
  \caption{Spacetime diagram of the temperature evolution for the hot
    nonrotating star in the simulation with the highest resolution. Also
    reported with coloured solid lines are the worldlines of the
    neutrinospheres in the three species considered here, while a white
    dashed line marks the worldline of the putative stellar surface,
    which is set to be where $\rho=10^{12}\mathrm{g\,cm}^{-3}$.}
  \label{fig:TOV_nu_radii}
\end{figure}

The oscillating behaviour of the unfiltered convergence order is most
likely due to the presence of additional, high-frequency oscillation
modes that are captured in the star evolved with the highest
resolution. Since the short timescale variability of the neutrino
luminosity is mostly dictated by the expansion/contraction of the
neutrinosphere, that is defined as the location where the optical depth
$\tau_{\nu_*}=1$ and which itself is regulated by the stellar
oscillations~\citep{Galeazzi2013, Perego2014}, the luminosities from
simulations with higher resolutions inevitably contain higher frequency
components, which add up when the errors are computed. Indeed, an
analogous oscillatory behaviour can also be observed when looking at the
locations of the neutrinospheres as reported in
Fig.~\ref{fig:TOV_nu_radii}. In particular, the figure offers a spacetime
diagram of the temperature evolution for the simulation at the highest
simulation. Also reported with coloured solid lines are the worldlines of
the neutrinospheres in the three species, while the white dashed line is
the worldline of the putative stellar surface, which we set to be where
the rest-mass density reaches a specific value, \ie
$\rho=10^{12}\,\mathrm{g\,cm}^{-3}$.

The challenging nature of this self-convergence test is clear also when
considering the convergence order computed on the pointwise values of the
central rest-mass density and lapse function reported in
Fig.~\ref{fig:TOV_conv} with coloured solid lines. Also for these
quantities, that are genuinely part of the vectors of numerically evolved
variables, a degradation of the expected convergence rate, \ie an order
of three~\citep{Most2019b}, is measured. This is not particularly
surprising and is due to the presence of the stellar surface, where small
shocks inevitably appear and lead to a degradation of the convergence
order~\citep[see][for a discussion]{Radice2012a, Radice2013c}. On the
other hand, the convergence order for the neutrino luminosities reported
in Fig.~\ref{fig:TOV_conv} is of the order of $1-1.5$, which is smaller
than expected second order implemented in \FMONEA, but aligned with what
already observed for the bulk spacetime and hydrodynamical variables. We
also note that the convergence order is somewhat different for the
various neutrino species. Since the radiative-transfer equations solved
are independent of the neutrino species, the reason for different
convergence behaviours should be sought in the different hydrodynamic and
thermodynamic conditions experienced by the different neutrino
species. More specifically, the different convergence behaviours follow
from the various depths within the star where the various neutrinos
decouple from the fluid and become free streaming. As can be clearly seen
in Fig.~\ref{fig:TOV_nu_radii} through the worldlines of the
neutrinospheres, heavy lepton neutrinos systematically decouple deeper
within the star than other neutrino species, whereas electron neutrinos
are strongly coupled with matter essentially up to the stellar
surface. Because the position of the latter is in general in a
lower-order convergent regime (large gradients and small shocks develop
near the surface), it is natural to expect that also electron neutrino
luminosities are those with the lowest convergence order. Moreover, the
oscillatory behaviour of the convergence order in Fig.~\ref{fig:TOV_conv}
is not the directly related to the eigenfrequencies of the underlining
star, but rather produced by the slightly different oscillation periods
at different resolutions. Overall, the results presented in section
provide evidence that the solution of the full set of the Einstein,
GRMHD, and radiative-transfer equations leads to an overall convergence
order which $\lesssim 3$ for the hydrodynamical and spacetime variables,
and $\lesssim 2$ for quantities related to neutrinos.

\subsection{Head-on collision of two neutron stars}
\label{sec:HEADON}

The final test of the robustness of our code is also the most involved
but hopefully also the most useful as it can be used in the future as an
effective, comparatively inexpensive but all-round benchmark test for the
implementation of an M1 scheme under realistic conditions of spacetime
curvature but also of hydrodynamical and thermodynamical states. In
particular, we consider the head-on collision of two equal-mass neutron
stars described by the temperature dependent DD2
EOS~\citep{Hempel2009}, each having a mass of $0.91\,M_{\odot}$ and a
central density of $\rho_c=4.6\times10^{14}\,\mathrm{g/cm}^3$. To
maintain the setup as simple as possible, we do not seek a
constraint-satisfying initial solution but simply consider the linear
combination of the spacetimes corresponding to two stars in isolation and
with a boost of $|v_{\rm in}|=\pm0.02$ along the direction of the
collision. Given this setup, the resulting object is not expected to
produce a black hole but a remnant star~\citep{Rezzolla2013,
  Koeppel2019}. The grid is set to have a total extent of $x_{\rm max}=
256\,M_{\odot}$ containing five nested fixed refinement levels at
positions $x_{^{\rm RL}}=\{ 32, 64, 80, 100, 160 \}\, M_{\odot}$.  The
outermost box has an outer boundary at and is covered by $40\times 40
\times 20$ points leading to a finest spacing of $\Delta x_{5} =
310.08\,\mathrm{m}$. Reflection symmetry across the $z-$plane is employed
to save computational resources.

\begin{figure*}
  \includegraphics[width=\textwidth]{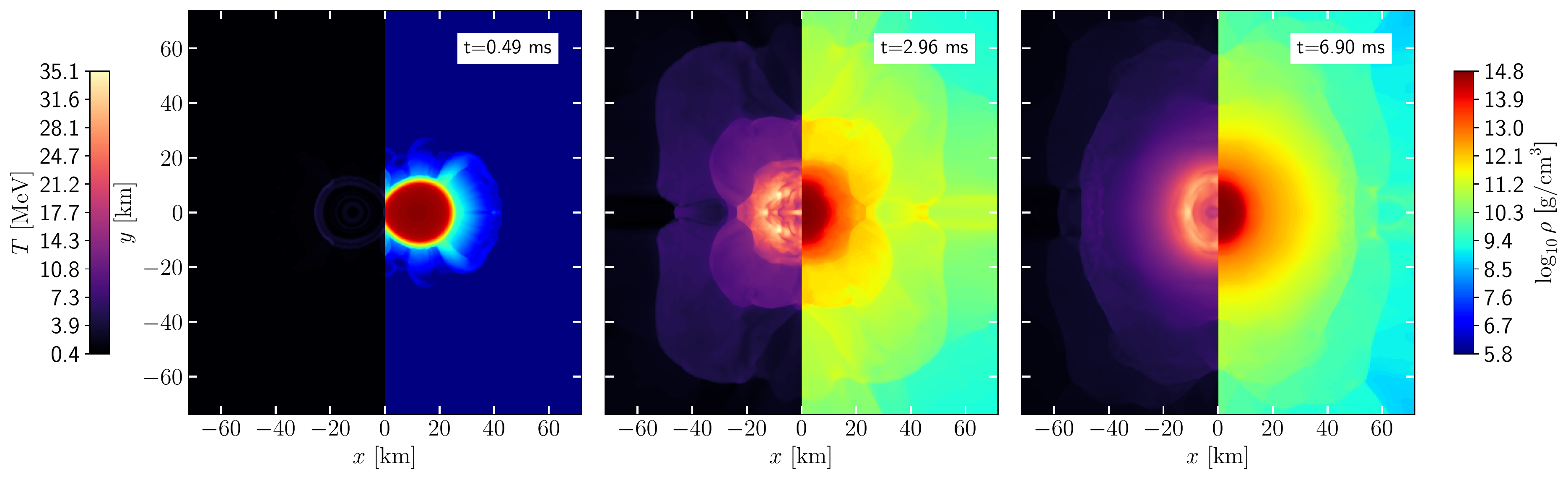}
  \caption{Representative snapshots of the head-on collision of two
    equal-mass neutron stars. All panels show the temperature (left half)
    and the rest-mass (right side). Note the appearance of a ring-like
    structure in the temperature of the relaxed post-collision remnant.}
  \label{fig:BNS_snap}
\end{figure*}

Figure~\ref{fig:BNS_snap} shows three representative snapshots of the
evolution of the colliding stars, reporting with colourcodes the
temperature distribution (left part of each panel) and the rest-mass
density distribution (right part of each panel). In the leftmost panel
the stars are about to collide and the temperature is very low everywhere
except near the stellar surfaces, where spurious heating takes place due
to interactions between the stars and the artificial atmosphere. The
central panel refers instead to an instant during the ``dynamical'' phase
where, the remnant star is undergoing violent oscillations and internal
shock heating. Finally, the rightmost
panel shows the state of the system at the end of the simulation,
consisting of a collision remnant that is secularly stable and that
reaches a roughly spherically symmetric state with a ring-like structure
in the temperature, as also encountered in the aftermath of BNS
mergers~\citep[see, \eg][]{Kastaun2014, Hanauske2016}. The origin of this
temperature distribution has to be found in the different manner in which
the thermal energy is produced and redistributed within the
remnant. While immediately after the violent collision of the two stars
(central panel of Fig.~\ref{fig:BNS_snap}) the temperature is
redistributed via strong shocks and hot material reaches the deepest
regions of the remnant's core, during the subsequent oscillations, the
temperature is redistributed mostly via conduction and the transport of
thermal energy of takes place along isopycnic levels and not across
them. As a result, the thermal energy will not be able to reach the dense
inner regions of the remnant, but will instead accumulate on an almost
spherical ring. It is also worth remarking that unlike a BNS merger,
where the stellar material is heated by the shearing of the two grazing
stars at merger, the heating here comes from the very strong shocks
produced by the collinear collision. As as a result, the temperature in
the remnant's core is almost a factor of ten larger than what is
typically be produced in a BNS merger remnant \citep[see
  \eg][]{Tootle2022}.

\begin{figure*}
  \includegraphics[width=\textwidth]{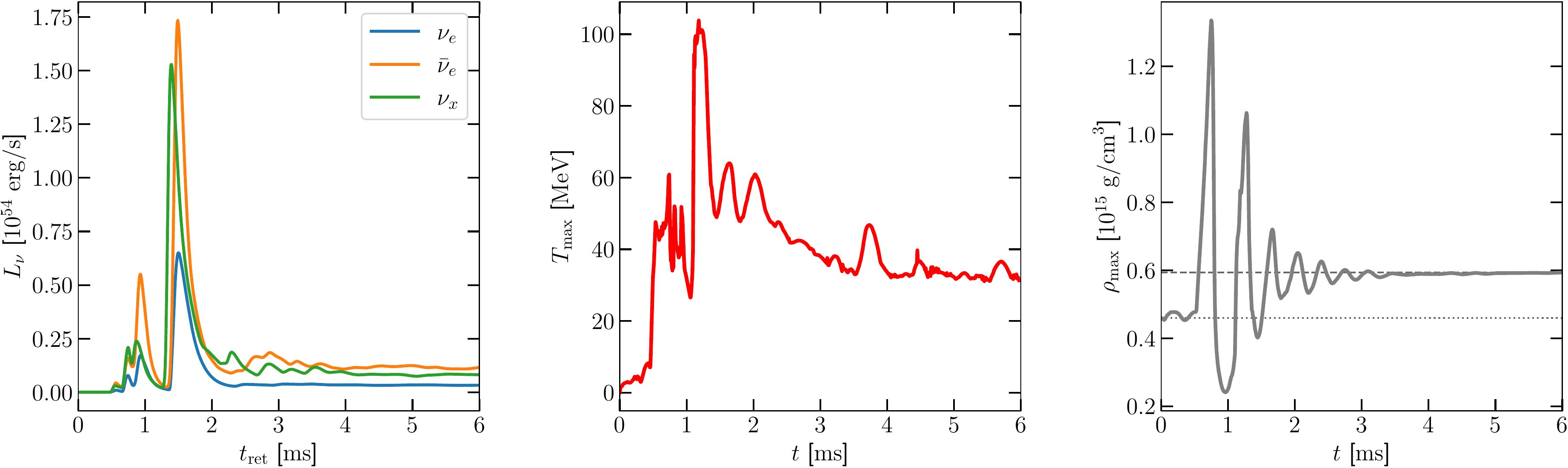}
  \caption{\textit{Left panel:} Neutrino luminosities as a function of
    the retarded time for the three neutrino species considered in the
    head-on collision of two neutron stars. Three strong peaks appear and
    are obviously correlated but not simultaneous and are produced by the
    initial violent oscillations of the remnant. As the oscillations are
    damped, the luminosities reach a very slow exponential decay due to
    the cooling of the remnant star. \textit{Middle panel}: Evolution of
    the maximum temperature, with the strongest peak occurring when the
    two disrupted cores bounce after the initial collision. Subsequently,
    the oscillations in the temperature are damped and the system enters
    a slow, neutrino-driven cooling stage. \textit{Right panel:}
    Evolution of the maximum rest-mass density, which clearly shows the
    behaviour typical of a damped oscillator. Note that the final central
    density is larger than the initial one (horizontal dotted line) but
    also slightly smaller than the one of a zero-temperature remnant
    with the same mass (horizontal dashed line), to which it
    progressively converges. }
  \label{fig:headon-luminosities}
\end{figure*}

Figure~\ref{fig:headon-luminosities} shows instead the evolution of the
neutrino luminosities for the three species (left panel), of the maximum
temperature (middle panel) and of the rest-mass density (right
panel). Note that the variations in the neutrino luminosities are not
quite simultaneous as they are produced by different neutrinospheres but
are strongly correlated with the corresponding variations in temperature
and rest-mass density which, in turn, are due to the violent oscillations
of the remnant system as it attains a new equilibrium. The strongest
peaks in the temperature are reached at $t=1.3\,{\rm ms}$, when the two
stellar cores bounce after the initial collision at $t=0.5\,{\rm ms}$,
after which the oscillations in the temperature are much smaller and the
system enters a slow, neutrino driven, cooling stage. Interestingly, we
record a slight delay between the oscillations in the rest-mass density
and those in the neutrino luminosity, which are likely due to the time
needed by the neutrinos to stream out of the remnant. After about
$3\,{\rm ms}$ most of the oscillations have been damped and the evolution
enters a quasi-stationary phase in which the the neutrino luminosities
reach almost constant values and the temperature experiences very slow
exponential decay due to cooling of the remnant star. Finally, note in
this new equilibrium the central rest-mass density of the remnant
stabilises around a value that is larger than the initial one (horizontal
dotted line) but also slightly smaller than the one of a
zero-temperature remnant with the same mass (horizontal dashed
line). This is due to the additional internal energy gained through the
collision that reduces the central density; as the remnant cools down,
the zero-temperature solution is progressively reached. The most
important data corresponding to this simulation, in terms of spacetime,
hydrodynamical, and thermodynamical quantities, will be made accessible
freely online for comparison with future M1-scheme implementations.

\section{Conclusions}

We have presented \FMONEA, a new implementation of a two-moment
radiative-transfer scheme for neutrino transport in numerical-relativity
simulations of neutron stars. \FMONEA incorporates many recent
algorithmic developments in radiation transport that contribute to its
stability and accuracy. In what follows we list briefly those steps in
the implementations that should be followed and the pitfalls that, on the
contrary, should be avoided.

\begin{itemize}
\item The collisional source terms of
  the M1 system become extremely stiff when the fluid velocity is very
  large. In these cases, resorting to simpler linearized expressions of
  the source terms is not advisable as it invariably leads to inaccurate
  or unstable evolutions of the system. On the other hand, the inversion
  of the full set of nonlinear equations (Eqs~\eqref{eq:implicit_eq}) together
  with implicit-explicit time-stepping, albeit more involved, has shown to lead to
  stable and accurate solutions.

\item Some radiative-transfer codes iterate the full state vector of the
  MHD and radiation variables to find the correct value for the implicit
  sources~\citep{MelonFuksman2019}. While mathematically correct, this
  approach is computationally unfeasible in simulations of neutron stars,
  where the conservative-to-primitive inversion procedure contributes to
  most of the computational costs. To circumvent this problem, the fluid
  state can be kept constant during the iterations of the nonlinear
  root-finding solver for the reaction source terms (which implies that
  the reaction rates are also kept fixed during said iteration). This
  approach can lead to oscillations near beta equilibrium that can
  amplify and lead to unstable evolutions near the stellar surface or the
  black-hole's apparent horizon. In these cases, fixing the reaction
  rates with the beta-equilibrated values of $T$ and $Y_e$
  [Eq.~\eqref{eq:beta_eq_fix}] has proven to be essential to obtain
  stable evolutions.

\item When computing the numerical fluxes at cell interfaces it is
  important that the scheme is asymptotically preserving: \ie~that the
  optically-thick limit the fluxes are those of a diffusion-type
  equation. The formulation employed by \FMONEA [see
  Eqs.~\eqref{eq:F_flux_mod} and \eqref{eq:HLLE_E_mod}] is asymptotically
  preserving and has two desirable features:

  \subitem-- it reduces to the three-state HLLE Riemann solver with the
  correct eigenspeeds of the M1 system in the low Peclet-number limit, so
  that the causal structure of the underlying partial differential
  equation is properly captured by the scheme.

  \subitem-- it circumvents the need of explicitly replacing the
  numerical flux with the correct asymptotic form, thus avoiding
  potential problems with the energy density in the fluid frame, as well
  as with the velocity-dependent terms.

\end{itemize}

Following these strategies, we have reported the results of a number of
tests for the validation of the solution of the radiative-transfer
equations, all of which \FMONEA passes successfully. In addition to these
standard tests in flat spacetimes, we also consider the validation of the
code across non-trivial scenarios of curved spacetimes such as that
involving the comparison with energy dependent but spherically symmetric
radiative-transfer calculations of core-collapse supernovae. Finally,
we introduced two novel tests which involve the solution of the full set
of the Einstein, GRMHD and radiative-transfer equations. These are the
study of the neutrino emission from an isolated hot neutron star and from
the head-on collision of two equal-mass neutron stars. While in the first
test we show that \FMONEA is able to provide solutions in a convergent
regime, the second test is particularly useful since the most important
data in terms of spacetime, hydrodynamical, and thermodynamical
quantities, is freely available and can be used as a benchmark test in
new implementations of the M1 schemes.

\section*{Acknowledgements}

We thank for insightful discussions and input E. R. Most, L. R. Weih, H. Ng,
R. De Pietri and K. Topolski. Partial funding comes from the State of Hesse
within the Research Cluster ELEMENTS (Project ID 500/10.006), by the ERC Advanced
Grant ``JETSET: Launching, propagation and emission of relativistic jets
from binary mergers and across mass scales'' (Grant No. 884631). LR
acknowledges the Walter Greiner Gesellschaft zur F\"orderung der
physikalischen Grundlagenforschung e.V. through the Carl W. Fueck
Laureatus Chair. The calculations were performed on the local ITP
Supercomputing Clusters Iboga and Calea and on HPE Apollo HAWK at the
High Performance Computing Center Stuttgart (HLRS) under the grant
BNSMIC.

\section*{Data Availability}

The data underlying the head-on collision test is publicly available and
can be accessed at the gitlab repository hosted at this
\href{https://gitlab.itp.uni-frankfurt.de/relastro_public/M1_data/m1-head-on-test-data}{link}.



\bibliographystyle{mnras}
\bibliography{aeireferences} 

\end{thebibliography}



\appendix

\section{Commonly employed weak rates}
\label{sec:appendix}

In this section we collect expressions for the weak rates employed in
\FMONEA for coupling radiation to the astrophysical plasma (see
Tab.~\ref{tab:reactions} for a detailed list of the processes we
include). All of these expressions can be found scattered in the
literature, but listing them here is useful for the sake of presenting a
complete account of all the necessary ingredients for constructing a grey
M1 scheme to be used in simulations of neutron star physics. Most of the
rates in this section come from~\citet{Ruffert96b, Bruenn85,
  Rosswog:2003b}.  Throughout this section and as is customary when
discussing weak-interaction rates, the rest-mass density $\rho$ will have
the units of $[\mathrm{g\,cm}^{-3}]$, whereas temperatures will be
expressed in $[\mathrm{MeV}]$ and factors of $c$ will be shown
explicitly. Hence, in these units, the opacities $\kappa$ correspond to
inverse mean-free-paths of neutrinos and have units of
$[\mathrm{cm}^{-1}]$, the energy production rates $Q^{j=1}$ have units of
$[\mathrm{MeV} \, \mathrm{s}^{-1} \, \mathrm{cm}^{-3}]$, and the
number-production rates have units of $[\mathrm{s}^{-1}\,
  \mathrm{cm}^{-3}]$. Furthermore, for the sake of compactness, in this
section we will indicate emissivities  by $Q^{j}$ with $j=1$ indicating
an energy rate and $j=0$ a number rate. Analogously, scattering (absorption)
opacities will be denoted by $\kappa_{s}^{j}$ ($\kappa_a^j$). We define the fugacity of
particle species $i$ as $\eta_i:=\mu_i/T$, where the neutrino chemical potentials $\mu_i$
are computed according to Eqs.~\eqref{eq:nue_mu_eq} and
\eqref{eq:nux_mu_eq} with a leakage-type correction factor of the form
$(1-\exp(-\tau_i))$, with $\tau$ the optical depth. On the other hand,
the chemical potentials of all other particles are extracted from the EOS
table.

Following~\citet{Ruffert96b}, we define the averaged absorption rate due
to absorption of electron neutrinos onto neutrons as
\begin{align}
  \label{eq:ka_nue}
	\kappa^{\,j}_{a,\nu_e} :=& \frac{1 + 3\alpha^2}{4}\, \sigma_0\,
        \xi_{np}\,\, \langle 1 - f(\epsilon_{e^-}; T, \eta_{e}) \rangle
        \,\times \notag \\ & \left( \frac{T}{m_{e} c^2} \right)^2\,
        \frac{\mathcal{F}_{4+j}(\eta_{\nu_e})}{\mathcal{F}_{2+j}(\eta_{\nu_e})}\,,
\end{align}
and for absorption of electron anti-neutrinos onto protons
\begin{align}
  \label{eq:ka_nua}
	\kappa^{\,j}_{a,\bar{\nu}_e} :=& \frac{1 + 3\alpha^2}{4}\,
        \sigma_0\, \xi_{pn}\,\, \langle 1 - f(\epsilon_{e^+}; T,
        -\eta_{e}) \rangle \,\times \notag \\ & \left( \frac{T}{m_{e}
          c^2} \right)^2\,
        \frac{\mathcal{F}_{4+j}(\eta_{\bar{\nu}_e})}{\mathcal{F}_{2+j}(\eta_{\bar{\nu}_e})}\,,
\end{align}
where $T$ is the matter temperature, $\alpha\simeq1.25$, and
$\sigma_0=1.76\times 10^{-44} \mathrm{cm}^2$. The coefficients $\xi_{pn}$
and $\xi_{np}$ are instead defined following~\citet{Bruenn85} as
\begin{align}
  \label{eq:xi_pn_np}
	\xi_{np} &:= \mathcal{A}\rho \frac{Y_p -
          Y_n}{e^{\eta_p-\eta_n}-1} = \mathcal{A}\rho Y_{np}\,,
        \\ \xi_{pn} &:= \mathcal{A}\rho \frac{Y_n -
          Y_p}{e^{\eta_n-\eta_p}-1} = \mathcal{A}\rho Y_{pn}\,,
\end{align}
where $\mathcal{A}\rho$ is the baryon number density and $Y_{np}, Y_{pn}$
are number the fractions corrected for nucleon phase-space Pauli
blocking. It is worth noting that with the assumption of charge
neutrality ($Y_p=Y_e$) and in simple neutron-proton-electron ($npe$)
matter where the presence of muons and other exotic degrees of freedom is
neglected, the neutron and proton fractions are simply given by $Y_n = 1
- Y_p$.

Fermi blocking for fermions is computed assuming that the energies of the
absorbed and emitted (anti-)leptons are the same and is approximated
as~\citep{Ruffert96b}
\begin{equation}
  \label{eq:fermi_blocking}
  \langle 1 - f(\epsilon_i; T, \eta_i) \rangle \simeq \Big\{ 1 +
  e^{-\left(\frac{\langle \epsilon_i \rangle}{T} - \eta_i \right)}
  \Big\}^{-1}\,,
\end{equation}
with the average energy taken to be
\begin{align}
  \langle \epsilon_{e^-} \rangle =
  T\frac{\mathcal{F}_{5}(\eta_{\nu_e})}{\mathcal{F}_{4}(\eta_{\nu_e})}\,,\qquad \langle
  \epsilon_{e^+} \rangle =
  T\frac{\mathcal{F}_{5}(\eta_{\bar{\nu}_e})}{\mathcal{F}_{4}(\eta_{\bar{\nu}_e})}\,.
\end{align}
The spectrally averaged opacity of neutrinos to scattering onto free
nucleons is given by~\citep{Ruffert96b}
\begin{equation}
  \label{eq:ks_1}
	\kappa^{\,j}_{s, \nu_i}(N) = \mathscr{C}_N\, \sigma_0 \, \xi_{NN} \, \left(
        \frac{T}{m_e c^2} \right)^2 \,\,
        \frac{\mathcal{F}_{4+j}(\eta_{\nu_i})}{\mathcal{F}_{2+j}(\eta_{\nu_i})}\,,
\end{equation}
where $N=n~(p)$ for neutrons (protons) and $\mathscr{C}_N$ is defined as 
\begin{equation}
  \label{eq:Cn}
	\mathscr{C}_{n} := \frac{1+5\alpha^2}{24}\,, 
\end{equation}
for scattering onto neutrons and
\begin{equation}
  \label{eq:Cp}
	\mathscr{C}_{p} = \frac{4(\mathscr{C}_V-1)^2+5\alpha^2}{24}\,, 
\end{equation}
with $\mathscr{C}_V=1/2+2\sin^2(\theta_W)$ and $\sin(\theta_W)\simeq 0.23$ for
scattering on protons. The nucleon fraction $\xi_{NN}$ is defined as
\begin{equation}
  \label{eq:xiNN}
	\xi_{NN} := \mathcal{A}\rho \,\,
        \frac{Y_N}{1+\frac{2}{3}\max(\eta_N, 0)} = \mathcal{A} \,\, \rho
        Y_{NN}\,,
\end{equation}
where $Y_{NN}$ is an interpolation between the particle fraction $Y_N$ in
the non-degenerate limit and $3Y_N/2\eta_{N}$ in the completely
degenerate one~\citep{Bruenn85}. Similarly, we include the effect of
scattering onto nuclei of mass number $A$ as
\begin{align}
  \label{eq:ks_2}
	\kappa^{\,j}_{s, \nu_i}(A) =& \frac{1}{6} A^2 \Big[ \mathscr{C}_A - 1
          +\frac{Z}{A}(2-\mathscr{C}_A - \mathscr{C}_V) \Big]^2 \,\times \notag \\ & \sigma_0
        \,n_A\,\, \left( \frac{T}{m_e c^2} \right)^2 \,
        \frac{\mathcal{F}_{4+j}(\eta_{\nu_i})}{\mathcal{F}_{2+j}(\eta_{\nu_i})}\,,
\end{align}
where, $\mathscr{C}_A=1/2$, while $Z$ is the charge number of the nucleus and $n_A$
its number density.

The emission of neutrinos through direct and inverse $\beta$ processes
represents the dominant sources for electron-flavour neutrinos and the
spectrally averaged production rates are~\citep{Ruffert96b}\footnote{Note
that we are neglecting the nucleon mass difference in all these
production rates, but the complete expressions have been reported
by~\citet{Ardevol2018}.}
\begin{align}
  \label{eq:Q_charged_current_nue}
	Q^j_{\nu_e, \beta} =& \frac{1+3\alpha^2}{8} \, \frac{\sigma_0
          c}{m_e c^2} \xi_{pn} \, \langle 1 - f(\epsilon^\beta_{\nu_e}; T
        \eta_{\nu_e}) \rangle \,\times \notag \\ 
	& \, \frac{8\pi}{(hc)^3}\, T^{5+j}\,\mathcal{F}_{4+j}(\eta_{e^-})\,, 
\end{align}
while for anti-neutrinos we have
\begin{align}
  \label{eq:Q_charged_current_anue}
  Q^j_{\bar{\nu}_e, \beta} =& \frac{1+3\alpha^2}{8} \, \frac{\sigma_0
    c}{m_e c^2} \xi_{pn} \, \langle 1 - f(\epsilon^\beta_{\bar{\nu}_e}; T
  \eta_{\bar{\nu}_e}) \rangle \,\times \notag \\ & \, \frac{8\pi}{(hc)^3}\,
  T^{5+j} \, \mathcal{F}_{4+j}(-\eta_{e^-})\,,
\end{align}
where we again compute blocking factors for fermions as we did for
absorption opacities [Eq.~\eqref{eq:fermi_blocking}] and with
\begin{align}
  \label{eq:Qcc_blocking_fermi}
  \langle \epsilon^\beta_{\nu_e} \rangle = T
  \frac{\mathcal{F}_5(\eta_{e^-})}{\mathcal{F}_4(\eta_{e^-})}\,,\qquad \langle \epsilon^\beta_{\bar{\nu}_e} \rangle 
  = T \frac{\mathcal{F}_5(-\eta_{e^-})}{\mathcal{F}_4(-\eta_{e^-})}\,.
\end{align}

The production rates for electron-positron pair annihilation relative to
electron-flavour neutrinos are given by
\begin{align}
  \label{eq:ee_Q}
	Q^j_{\nu_e,\bar{\nu}_e, ee} =& \frac{(\mathscr{C}_1+\mathscr{C}_2)_e}{72}
        \frac{\sigma_0c}{(m_e c^2)^2} \,\times \notag \\ & \langle 1 -
        f(\epsilon^{ee}_{e} ; T, \eta_{\nu_e}) \rangle \, \langle 1 -
        f(\epsilon^{ee}_{e} ; T, \eta_{\bar{\nu}_e}) \rangle \,\, \left(
        \frac{8 \pi}{(hc)^3} \right)^2 \,\times \notag \\ & T^{8+j} \big[
          \mathcal{F}_{3+j}(\eta_e) \mathcal{F}_{3}(-\eta_e) +
          \mathcal{F}_{3+j}(-\eta_e) \mathcal{F}_{3}(\eta_e) \big]\,,
\end{align}
where we define the constant $(\mathscr{C}_1+\mathscr{C}_2)_e := (\mathscr{C}_V-\mathscr{C}_A)^2 + (\mathscr{C}_V + \mathscr{C}_A)^2$
and the mean neutrino energy in the Pauli blocking term is taken to be
\begin{equation}
  \label{eq:eps_ee_1}
	\langle \epsilon^{ee}_{e} \rangle = T \left(
        \frac{1}{2}\frac{\mathcal{F}_4(\eta_{e})}{\mathcal{F}_3(\eta_{e})}
        +
        \frac{1}{2}\frac{\mathcal{F}_4(-\eta_{e})}{\mathcal{F}_3(-\eta_{e})}
        \right)\,.
\end{equation}
On the other hand, for the pair production rate effective heavy lepton
neutrino species is given by 
\begin{align}
  \label{eq:ee_Qx}
	Q^j_{\nu_x, ee} =& \frac{(\mathscr{C}_1+\mathscr{C}_2)_x}{18} \frac{\sigma_0c}{(m_e
          c^2)^2} \,\times \notag \\ & (\langle 1 - f(\epsilon^{ee}_{e} ;
        T, \eta_{\nu_x}) \rangle)^2 \,\, \left( \frac{8 \pi}{(hc)^3}
        \right)^2 \,\times \notag \\ & T^{8+j} \big[
          \mathcal{F}_{3+j}(\eta_e) \mathcal{F}_{3}(-\eta_e) +
          \mathcal{F}_{3+j}(-\eta_e) \mathcal{F}_{3}(\eta_e) \big]\,,
\end{align}
and $(\mathscr{C}_1+\mathscr{C}_2)_x := (\mathscr{C}_V-\mathscr{C}_A)^2 + (\mathscr{C}_V + \mathscr{C}_A - 2)^2$. 
The decay of transverse plasmons also contributes to the emission of all
neutrino species and is expressed as
\begin{align}
  \label{eq:gamma_Q_1}
	Q^j_{\nu_e,\bar{\nu}_e, \gamma} \simeq& \frac{\pi^3}{3\alpha^*}
        \mathscr{C}_V^2 \frac{\sigma_0c}{(m_e\,c^2)^2} \frac{T^8}{(hc)^6}
        \gamma^6e^{-\gamma}(1+\gamma) \,\times \notag \\ & \langle 1 -
        f(\epsilon^{\gamma}_{e} ; T, \eta_{\nu_e}) \rangle \, \langle 1 -
        f(\epsilon^{\gamma}_{e} ; T, \eta_{\bar{\nu}_e}) \rangle \,\times
        \notag \\ & \left\{ \frac{1}{2} T \left( 2 +
        \frac{\gamma^2}{1+\gamma} \right)\right\}^j\,,
\end{align}
for electron-flavour neutrinos and for the other species as
\begin{align}
  \label{eq:gamma_Q_2}
	Q^j_{\nu_x, \gamma} &\simeq \frac{\pi^3}{3\alpha^*} \left(\mathscr{C}_V-1\right)^2
        \frac{\sigma_0c}{(m_e\,c^2)^2} \frac{T^8}{(hc)^6}
        \gamma^6e^{-\gamma}(1+\gamma) \,\times \notag \\ & \left(\langle 1 -
        f(\epsilon^{\gamma}_{e} ; T, \eta_{\nu_x}) \rangle\right)^2 \, \left\{
        \frac{1}{2} T \left( 2 + \frac{\gamma^2}{ 1+\gamma }
        \right)\right\}^j\,,
\end{align}
where $\alpha_*=1/137.036$ is the fine-structure constant,
$\gamma=5.565\times10^{-2}\sqrt{1/3(\pi^2+3\eta_{e}^2)}$, and the mean
energy of the produced neutrinos in plasmon decay we take to be
\begin{equation}
  \label{eq:eps_ee_2}
	\langle \epsilon^{\gamma}_{e} \rangle = \frac{1}{2} T \left( 2 +
        \frac{\gamma^2}{1+\gamma} \right) \,.
\end{equation}

Finally, we include production of heavy lepton neutrinos via
nucleon-nucleon bremsstrahlung as~\citep{Burrows2006b}
\begin{equation}
  \label{eq:brems_Q}
	Q^{j=1}_{\nu_x, {\rm brems}} =  2.08 \times 10^2 \,\, \zeta \, \rho^2
        \, \left( Y_n^2 + Y_p^2 + \frac{28}{3}Y_n\,Y_p \right) \,\, T^{5.5}\,,
\end{equation}
where we set $\zeta=0.5$~\citep{Burrows2006b}. It is important to note
that the reported rate is for all four neutrino species combined. We then
follow~\citet{Ardevol2018} and set
\begin{equation}
  \label{eq:brems_R}
	Q^{j=0}_{\nu_x, {\rm brems}} = \frac{Q^{j=1}_{\nu_x, {\rm brems}}}{3T}\,.
\end{equation}

\bsp	
\label{lastpage}
\end{document}